\documentclass{emulateapj}
\shorttitle{QLF based on Y-band selected sample}
\shortauthors{Yang et al.}

\begin{document}


\title{Deep CFHT $Y$-band Imaging of VVDS-F22 Field: II. Quasar Selection and Quasar Luminosity Function}


\author{Jinyi Yang\altaffilmark{1,2}, Xue-Bing Wu\altaffilmark{1,2}, Dezi Liu \altaffilmark{1}, Xiaohui Fan\altaffilmark{3,2}, Qian Yang\altaffilmark{1,2}, Feige Wang\altaffilmark{1,2}, Ian D. McGreer\altaffilmark{3}, Zuhui Fan\altaffilmark{1}, Shuo Yuan\altaffilmark{1}, Huanyuan Shan\altaffilmark{4,5}}

\altaffiltext{1}{Department of Astronomy, School of Physics, Peking University, Beijing 100871, China}
\altaffiltext{2}{Kavli Institute for Astronomy and Astrophysics, Peking University, Beijing 100871, China}
\altaffiltext{3}{Steward Observatory, University of Arizona, 933 North Cherry Avenue, Tucson, AZ 85721, USA}
%\altaffiltext{4}{Laboratoire d'astrophysique (LASTRO), Ecole Polytechnique F\'{e}d\'{e}rale de Lausanne (EPFL), Observatoire de Sauverny, CH-1290 Versoix, Switzerland}
\altaffiltext{4}{Argelander-Institut f\"ur Astronomie, Auf dem H\"ugel 71, 53121 Bonn, Germany}
\altaffiltext{5}{Laboratoire d'astrophysique (LASTRO), Ecole Polytechnique F\'ed\'erale de Lausanne (EPFL), Observatoire de Sauverny, CH-1290 Versoix, Switzerland}



\begin{abstract}
We report the result of a faint quasar survey in a one square degree field. The aim is to test the $Y-K/g-z$ and $J-K/i-Y$ color selection criteria for quasars at faint magnitude, to obtain a complete sample of quasars based on deep optical and near-infrared color-color selection, and to measure the faint end of quasar luminosity function (QLF) over a wide redshift range. We carried out a quasar survey based on the $Y-K/g-z$ and $J-K/i-Y$ quasar selection criteria, using the deep $Y$-band data obtained from our CFHT/WIRCam $Y$-band images in a two-degree field within the F22 field of the VIMOS VLT deep survey, optical co-added data from Sloan Digital Sky Survey Stripe 82 and deep near-infrared data from the UKIDSS Deep Extragalactic Survey in the same field. We discovered 25 new quasars at $0.5 < z < 4.5$ and $i < 22.5$ mag within one square degree field. The survey significantly increases the number of faint quasars in this field, especially at $z \sim 2-3$. It confirms that our color selections are highly complete in a wide redshift range ($z < 4.5$), especially over the quasar number density peak at $z \sim 2-3$, even for faint quasars. Combining all previous known quasars and new discoveries, we construct a sample with 109 quasars, and measure the binned QLF and parametric QLF. Although the sample is small, our results agree with a pure luminosity evolution at lower redshift and luminosity evolution and density evolution model at redshift $z > 2.5$.

\end{abstract}


\keywords{galaxies: active - galaxies:high-redshift - quasars: general - quasars: emission lines}



\section{Introduction}
Quasar luminosity function (QLF) has been the most important tool to directly characterize the evolution of quasar number density with redshift and luminosity for a half century. The measurement of QLF highly depends on the sample of quasars. Quasar samples currently available are usually incomplete due to various problems (e.g. photometry depth, instruments limitation), especially in the quasar candidate selections \citep{richards02}. For example, both 2dF QSO Redshift Survey \citep{boyle00} and Sloan Digital Sky Survey (SDSS) \citep[e.g. DR7 quasar catalog; ][]{schneider10} adopted UV excess-based  technique to select $z < 2.2$ quasars. More recent surveys focusing on quasars at $z \sim 2-3$ and higher redshifts have improved quasar samples at these redshifts \citep{ross12, myers15}. The SDSS-III Baryon Oscillation Spectroscopic Survey \citep[BOSS,][]{dawson13, ross12} highly improved quasar selection at $z > 2.2$ and spectroscopically identified $\sim$ 170,000 new quasars at $2.1 < z < 3.5$ to the depth of $g < 22$.
The SDSS IV extended Baryon Oscillation Spectroscopic Survey \citep[eBOSS,][]{dawson15, myers15} adopting two approaches, one from the combination of likelihood-based optical selection with mid-IR-optical color cut, the other one from variability, aims at targeting more quasars at $z > 2.2$. While, the current $z \sim 2-3$ quasar surveys are still not highly complete. Because the selections, in color space, always need to avoid the whole region that is seriously contaminated by stars.

Based on the previously suggested $K$-band excess technique \citep{hewett06, maddox08, warren00}, \cite{wu-jia10} posed two new selection criteria involving both optical and near-infrared (NIR) colors for selecting quasars at $z<4$ and $z<5$, respectively. They found that quasars at redshift $z < 4$ could be separated from stars well in the $Y-K$ versus $g-z$ color-color diagram, while quasars at $z > 4$ begin to enter the loci of stars due to the shift of strong Ly$\alpha$ emission line. In this case, the $J-K$/$i-Y$ color-color diagram has been suggested as effective in separating quasars at $4 < z < 5$ from stars. \cite{wu-jia10} cross-matched SDSS DR7 quasar catalog with the UKIRT InfraRed Deep Sky Surveys (UKIDSS) \citep{casali07, hewett06, lawrence07} DR3 NIR photometric data to obtain a sample of 8498 quasars with both SDSS and UKIDSS photometry. They tested the $Y-K$/$g-z$ and $J-K$/$i-Y$ selection criteria with this quasar sample, and found that the $Y-K$/$g-z$ color cut could select 98.6\% of $z < 4$ known quasars, and $J-K$/$i-Y$ cut could recover 97.5\% of $z < 4$ quasars, and 99\% of $4 < z < 5$ quasars. Some spectroscopic observations carried out by \cite{wu10a, wu10b} and \cite{wu11} have also demonstrated the effectiveness of using the SDSS-UKIDSS optical/NIR colors to find SDSS missing quasars at $2.2 < z < 3.5$. Therefore, these selection criteria are expected to be helpful for the construction of a relatively more complete quasar sample at $z < 5$, especially at the range of $2 < z < 3$. In addition, \cite{wu-jia10} only tested their selection criteria with relative bright quasar sample. A fainter sample is necessary to test this selection method at faint end.

Therefore, to construct a more complete quasar sample at $z < 4$, especially at the faint end, we used these two selection criteria to select quasar candidates in a deep optical/NIR surveyed field. VVDS F22 is a wide field covering a sky area of 4 square degrees \citep{garilli08, lefevre13}. This field was mapped by the VIMOS VLT deep survey (VVDS) deep optical (U,B,V,R,I) and NIR ($J, K$, but restricted access) photometry, which reached the depth of I $\sim$ 25 mag and K $\sim$ 23 mag (AB). It was also covered by the SDSS stripe 82 \citep{annis14} deep optical photometry, CFHT Legacy Survey (CFHTLS), the UKIDSS - Large Area Survey (LAS)  \& Deep Extragalactic Survey (DXS), and NRAO Very Large Array Sky Survey (NVSS) \citep{condon98}. Since there is no existing deep Y-band photometry, we first obtained deep Y-band image in this field and then carried out a small field but deep quasar spectroscopic survey. VVDS spectroscopy has been done with a sampling rate of 22\% for sources with I$<$22.5, and has obtained the spectra of 11228 galaxies, 6748 stars and 167 quasars \citep{garilli08, lefevre13}. With new deep $Y$-band photometric data, we will be able to construct a more complete quasar sample. This sample will also enable the measurement of the faint end of QLF. 

In this paper, we report our work on a faint quasar survey based on our deep $Y$-band imaging in a square degree field of VVDS F22. We will describe the $Y$-band imaging and quasar candidates selection in Section 2. The spectroscopic observations, new discoveries and the construction of a quasar sample will be presented in Section 3. In Section 4 and 5, we will discuss the completeness of our survey and measure the QLF in a one square degree field. We adopt a $\Lambda$CDM cosmology with parameters $\Omega_{\Lambda}$ = 0.728, $\Omega_{m}$ = 0.272, $\Omega_{b}$ = 0.0456, and H$_{0}$ = 70 $km s^{-1} Mpc^{-1}$ \citep{komatsu09}. Photometric data from the Sloan Digital Sky Survey (SDSS) are in the SDSS photometric system\citep{lupton99}, which is almost identical to the AB system at bright magnitudes; photometric data from NIR surveys are in the Vega system. 

\section{Candidates selection}
\subsection{Deep $Y$-band photometry}
We obtained $Y$-band imaging of a two square degree field (see Fig.1) within F22 field of the VVDS using CFHT WIRCam in August, September and October, 2012. The WIRCam focal plane is made of a mosaic which includes four HAWAII2-RG detectors. The field of view of the full mosaic is $21.5' \times 21.5'$. We divided our two square degree field into 18 sub-fields, each of $20' \times 20'$, as shown in Figure 1. The final imaged area is about $128' \times 65'$, fully covering those 18 sub-fields. More details of $Y$-band imaging can be found in \cite{liu17}, including observation, data reduction, data release and $Y$-band related photometric redshift measurements.

The $Y$-band photometric data used in this paper for quasar survey was the earlier version (in 2013) of the finally published data (in 2017). The deep Y-band image was processed by SIMPLE-WIRCAM pipeline \cite{wang10} specified for CFHT/WIRCAM image analyses. The basic steps of CFHT/WIRCAM image reduction encoded in the pipeline contains flat-fielding, sky subtraction, cosmic ray removal, astrometric calibration and final image stacking. We used SExtractor \citep{bertin-arnouts96} and DAOPHOT for source extraction and photometry (aperture photometry by DAOPHOT). In total,  $\sim$110000 sources were detected within this field. The final Y-band photometric data has been published by \cite{liu17}, which was an improved version for photometric redshift measurements (e.g. PSF homogenization of different bands). The 5$\sigma$ magnitude limit of Y-band photometry is 22.25 mag in Vega magnitude system.

\begin{figure}%f1
\centering
\includegraphics[width=0.5\textwidth]{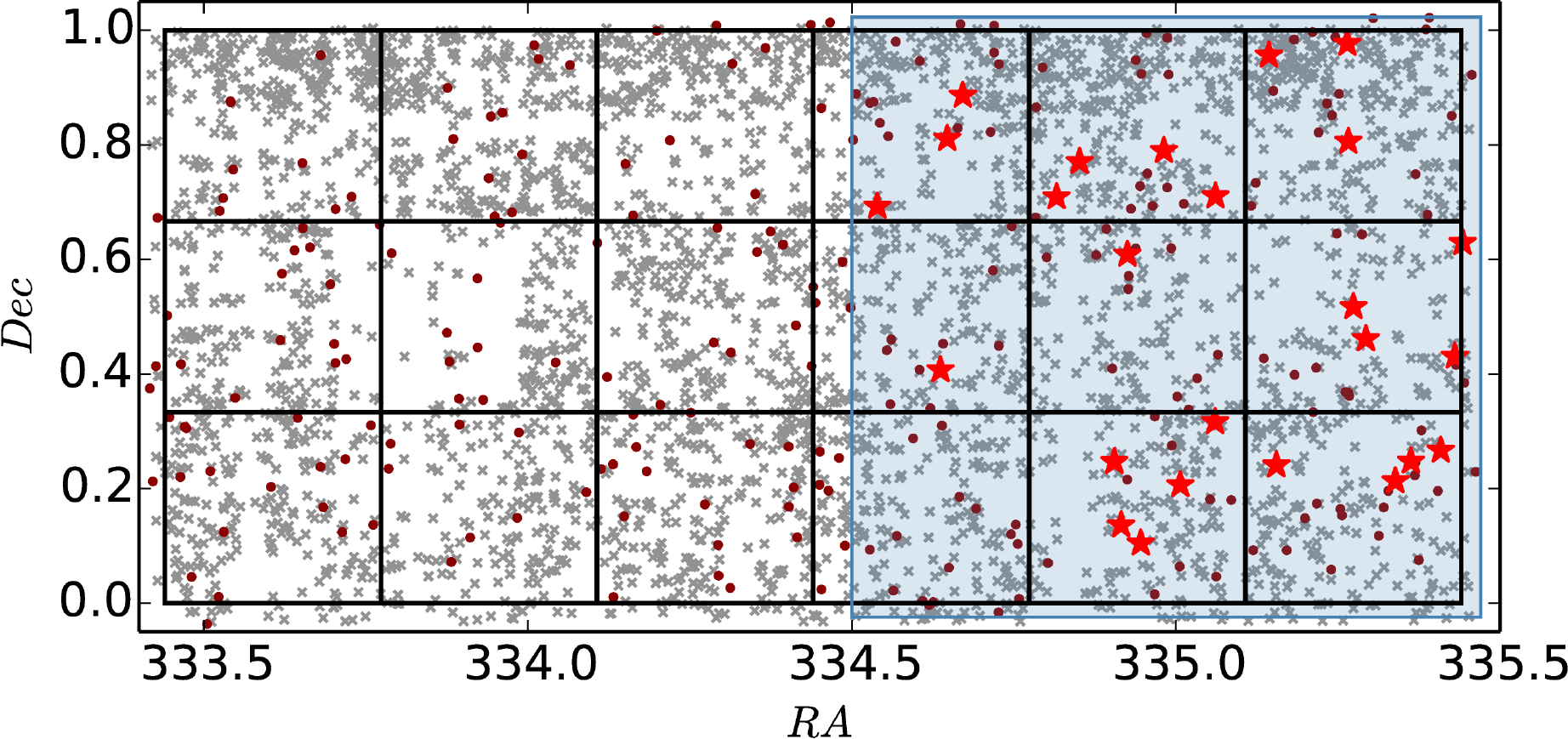}
\caption{Sky coverage of the deep $Y$-band image. The black frames represent 18 sub-fields, each of $20' \times 20'$. The final imaged area is about $128' \times 65'$. The dark red dots and grey crosses denote spectroscopically identified quasars and galaxies by previous works \citep[e.g. SDSS DR7, 9, 10, 12 quasar catalog and VVDS spectroscopy;][]{schneider10, paris12, paris14, paris17, garilli08, lefevre13}. The shaded region represents the area of our spectroscopy survey, and new quasars from our survey are marked as red stars.}
\label{fig1}
\end{figure}

\subsection{Quasar candidate selection}
\begin{figure*}%f2
\includegraphics[width=0.8\textwidth]{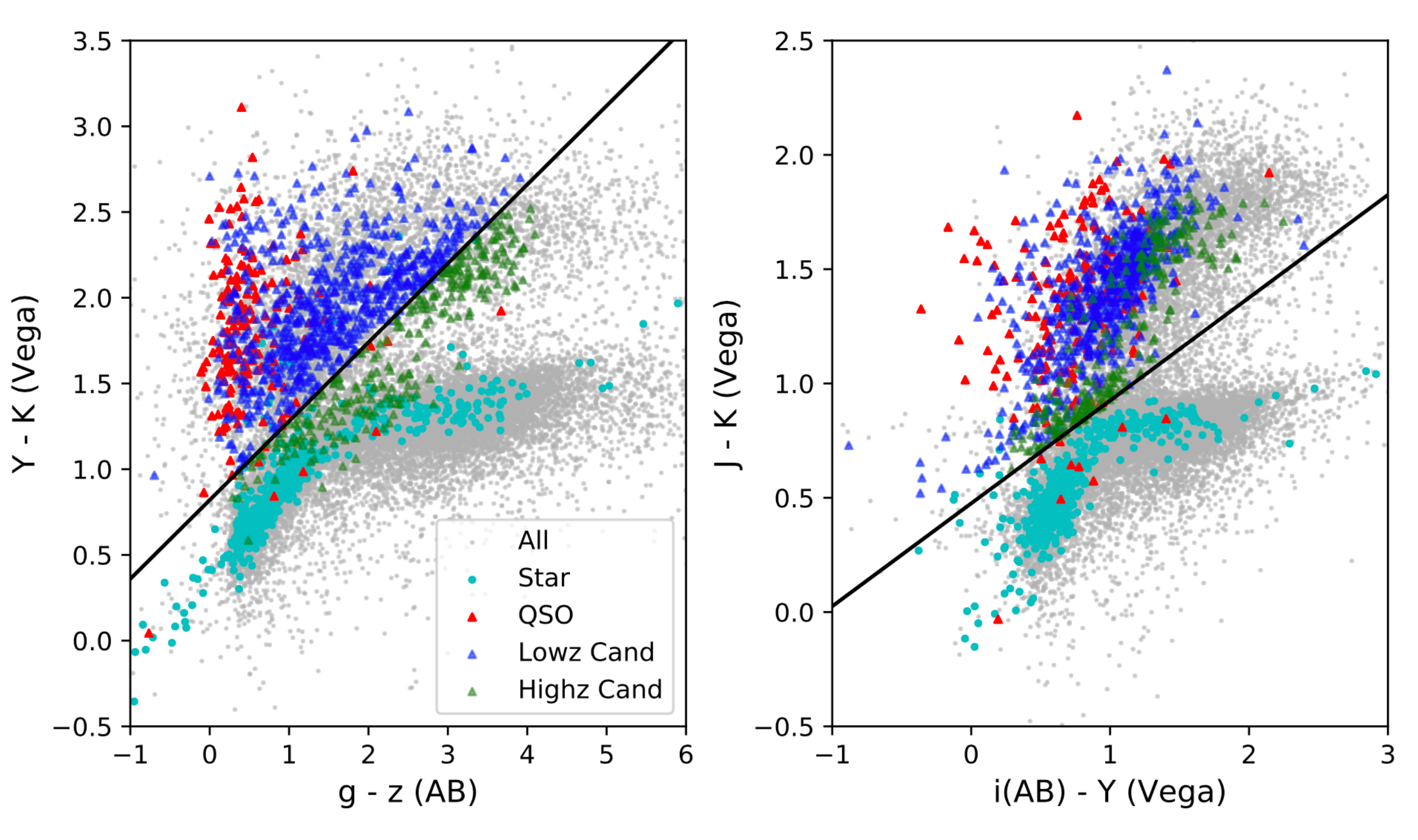}
\centering
\caption{ The $Y-K$/$g-z$ and $J-K$/$i-Y$ color-color diagrams and cuts used to select quasar candidates from all point sources. The grey points represent all sources within this field. The previously known quasars (red triangles), stars (cyan dots) are also plotted. Sources met $Y-K$/$g-z$ or $J-K$/$i-Y$ color-based selection criteria were denoted by blue triangles (low redshift quasar candidates) and green triangles (high redshift quasar candidates).}
\label{fig2}
\end{figure*}

We started our selection with the catalog of deep Y-band detected sources in the region of $333.40^\circ \le RA \le 335.47^\circ$ and $-0.05^\circ \lesssim Decl. \lesssim 1.024^\circ$. We cross matched (3") the Y-band source catalog with the co-added SDSS Stripe 82 catalog from \cite{annis14}\footnote{The final catalog was published in 2014. But we used the older version of catalog from Annis et al. (2011, arXiv: 1111.6619v2) when we selected candidates in 2013.} and UKIDSS UDXS DR9 catalog to obtain required photometric data in $g, i, z, Y, J$ and $K$ bands. We obtained a sample of $\sim$ 105,000 sources.

%point/extended separation
Since our selection focused on faint objects, galaxy contamination is serious. We use SDSS data for star/galaxy separation. Our two square degree field was covered by SDSS Stripe82 photometry, which is a 275 square degree region on the Celestial Equator in the Southern Galactic Cap. This region was imaged multiple times by SDSS $u, g, r, i, z$ bands during the fall seasons when the North Galactic Cap was not observable. After a co-addition of multiple epoch images, Stripe82 data reaches a depth more than 2 magnitude fainter than the SDSS main survey and a $\sim$ 1.1$''$ median seeing in $r$ band \citep{annis14}. The stringent point/extended separator used by Stripe 82 data release, $|r_{psf} -r_{model}|$ $\le$ 0.03, could significantly reduce the number of misclassified galaxies, comparing with the SDSS standard star/galaxy separation ($|r_{psf} -r_{model}|$ $\le$ 0.145). But when testing with a sample of previously known quasars, we found this separation would misclassify some faint or low redshift known quasars. So to improve the completeness of quasar selection, we need to relax the separation. We used a sample of Stripe 82 detected previously known quasars, stars and galaxies to draw a new cut, $|r_{psf} -r_{model}|$ $\le$ 0.1. This cut help us to selected $\sim$ 40,000 sources from the whole sample.

 %coloc-color selection
We selected quasar candidates from all point sources by using the $Y-K$/$g-z$ and $J-K$/$i-Y$ selection criteria. The selection criteria we used are listed below. We also limited the magnitude errors of $g, i, z$ bands to be smaller than 0.5 and image quality flags of $Y, J, K$ bands to be 0, which mean clean photometry. The J/Kflags (also named as jppErrBits and kppErrBits) were obtained from UKIDSS UDXS DR9. The J/Kflags is the post-processing error quality bit flags assigned in the Wide Field Camera Science Archive (WSA) curation procedure for survey data. A J/Kflags $>$ 0 represents bad image quality (e.g. blended, bad pixel(s) in default aperture, close to saturated). The Y-band image flag was set by our data reduction in the same way to UDXS data. Considering the flux limit of MMT spectroscopic observation and the limited exposure time, we select objects with $i$ band magnitude brighter than 22.5 mag. When we selected candidates for spectroscopy, we required that the object meets equations (1) - (4), and then either (5)\&(6) or (7)\&(8). Since the $Y-K$/$g-z$ cut focuses on quasars at $z < 4$ and $J-K$/$i-Y$ can recover most of quasars at $4 < z < 5.3$, we also marked candidates that met $J-K$/$i-Y$ selection but not $Y-K$/$g-z$ as high redshift quasar candidates. Here, the $i$ band magnitude used in the $J-K$/$i-Y$ color-color cut (eq. 7) has been converted to Vega magnitude by $i_{Vega}$ = $i_{AB}$ - 0.366. The color - color diagrams are plotted in Fig 2. As shown, the color cuts can recover most of previously known quasars. All optical data used for color-color cuts are corrected for Galactic extinction. After color-color selection, we restricted candidate sample to $\sim$ 1,300 sources.

\begin{equation}
i < 22.5;
\end{equation}
\begin{equation}
Yflag = 0;
\end{equation}
\begin{equation}
Jflags = 0;
\end{equation}
\begin{equation}
Kflags = 0;
\end{equation}
~and~
\begin{equation}
Y - K > 0.46(g - z) + 0.82
\end{equation}
\begin{equation}
err_g < 0.5 ~and~ err_z <0.5
\end{equation}
~or~
\begin{equation}
J - K > 0.45(i - Y) + 0.475
\end{equation}
\begin{equation}
err_i <0.5
\end{equation}

 %chi2 selection
We then used a $\chi^{2}$ estimations to further rejects star contaminations from the color selected candidate sample. The $\chi^{2}$ represents a $\chi^{2}$ fitting of each object's photometric data to the quasar color-z relations \citep{wu-jia10, yang17}. 
The quasar color-z relation was generated by using a sample of real quasars. We first calculated the mean colors at each redshift bin and rejected quasars with any color out of $3\sigma$ to the mean value. Then we constructed the color-z relation using remaining quasars. 
Following the method given by \cite{weinstein04}, we calculated the mean color vector $M_{i}$ and the covariance matrix $V_{i}$ in the $i$th redshift bin. For each candidate, based on its photometric data, we could get the magnitude error matrix $V_{0}$. Then we computed the $\chi^{2}$ value between the colors of candidate and color-z relation in the $i$th redshift bin: $\chi_{i}^{2}$ = ($X_{0} - M_{i})^{T}(V_{0}+V_{i})^{-1}(X_{0}-M_{i})$, where the vector $X_{0}$ represents the observed colors of a candidate \citep{weinstein04}. From this $\chi^{2}$ value, we could derive the probability that a candidate lay in the $i$th redshift bin. But here we only used the minimum $\chi^{2}$ value of a candidate for star/quasar separation. Photometric data in $u, g, r, i, z, Y, J$ and $K$ bands were involved in the $\chi^{2}$ calculation.

A smaller $\chi^{2}$ value means that the colors of this source are more similar to the quasar colors at a given redshift than star's colors. Thus this source is expected to have higher probability to be a quasar. We tested the fitting by using spectroscopic identified quasars and stars and found that the $\chi^{2}$ of most quasars were relatively smaller than that of stars. So we can define a $\chi^{2}$ cut to separate quasars and stars. Generally, a higher $\chi^{2}$ limit corresponds higher quasars selected fraction (high completeness) but also higher fraction of stars contamination (low efficiency). We finally used the limit of $\chi^{2}$ value less than 15 to separate quasar from star, which is a empirical cut generated based on the $\chi^{2}$ distribution of known quasars and stars. 

Based on the selections described above, after removing all previously known objects, we finally selected about 550 quasar candidates in the two square degree field. All quasar candidates were also divided into three parts and set as different ranks. Candidates with different ranks had different priorities to be targeted and observed. The first rank with highest priority included candidates with smaller $\chi^{2}$ value ($< 10$). The second rank represented candidates with larger $\chi^{2}$ value ($> 10$) but better morphology ($|r_{psf} -r_{model}|$ $\le$ 0.03), and remaining candidates were marked as rank 3. There are $\sim$ 290 rank 1 candidates, $\sim$ 100 rank 2 candidates and $\sim$ 160 rank 3 candidates.

\section{Results}
\subsection{Observations}

Our quasars candidates were observed by MMT/Hectospec \citep{fabricant05}. We divided our two square degree field into two 1 square degree fields to match the Hectospec focus plane. Considering the fibers density and the efficiency of fiber configuration ($\sim$ 70\% $-$80\%), we finally submitted our quasar candidates together with about 240 galaxy candidates which were related with another project. When we did fiber allocation, we chose a configuration that could target more candidates in the first rank. After fiber fitting, 280 quasar candidates were targeted by MMT Hectospec. Our spectroscopically identifications have been done on October 8, 9 and November 29, 2013 with MMT Hectospec in a one square degree field ($334.5^\circ < RA < 335.47^\circ$). In this one square degree field, 143 quasar candidates had been observed. Candidates in the other one square degree field were not observed due to the limited observing time. The effective exposure time for final identification is 3$\times$30 minutes in total with a average seeing $\sim$ 0.52".  

\subsection{New Quasars}
\begin{figure*}%f3
\includegraphics[width=0.9\textwidth]{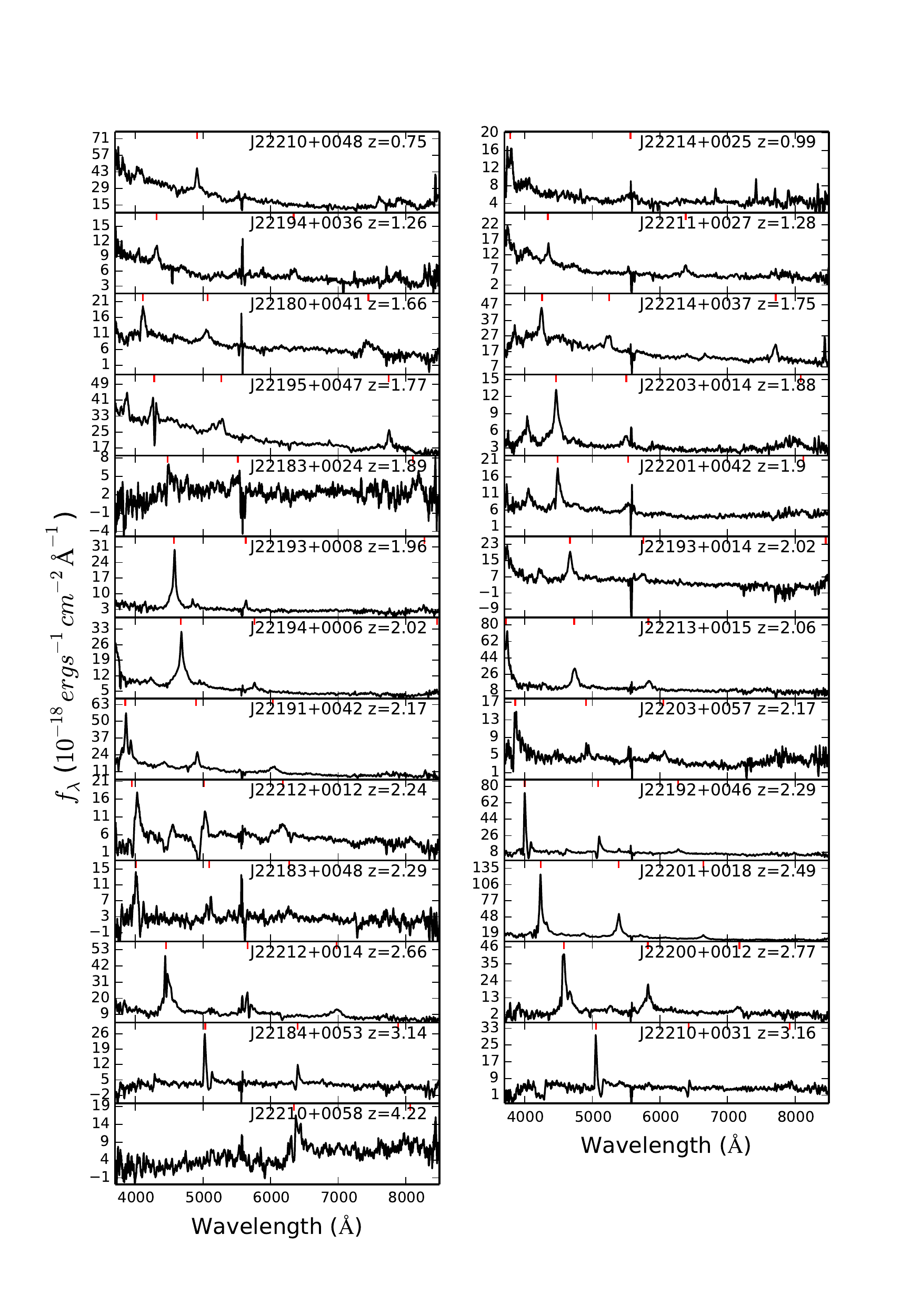}
\centering
\caption{The spectra of 25 newly discovered quasars in the one square degree field. They are smoothed with 15 pixel (1.21$\rm \AA$/pixel) box char. The red vertical lines show the $\rm Ly \alpha$, C\,{\sc iv} and Mg\,{\sc II} emission lines. All spectra are corrected for Galactic extinction using the \cite{cardelli89} Milky Way reddening law and E(B $-$ V) derived from the \cite{schlegel98} dust map.}
\label{fig3}
\end{figure*}

\begin{figure}%f4
\includegraphics[width=0.48\textwidth]{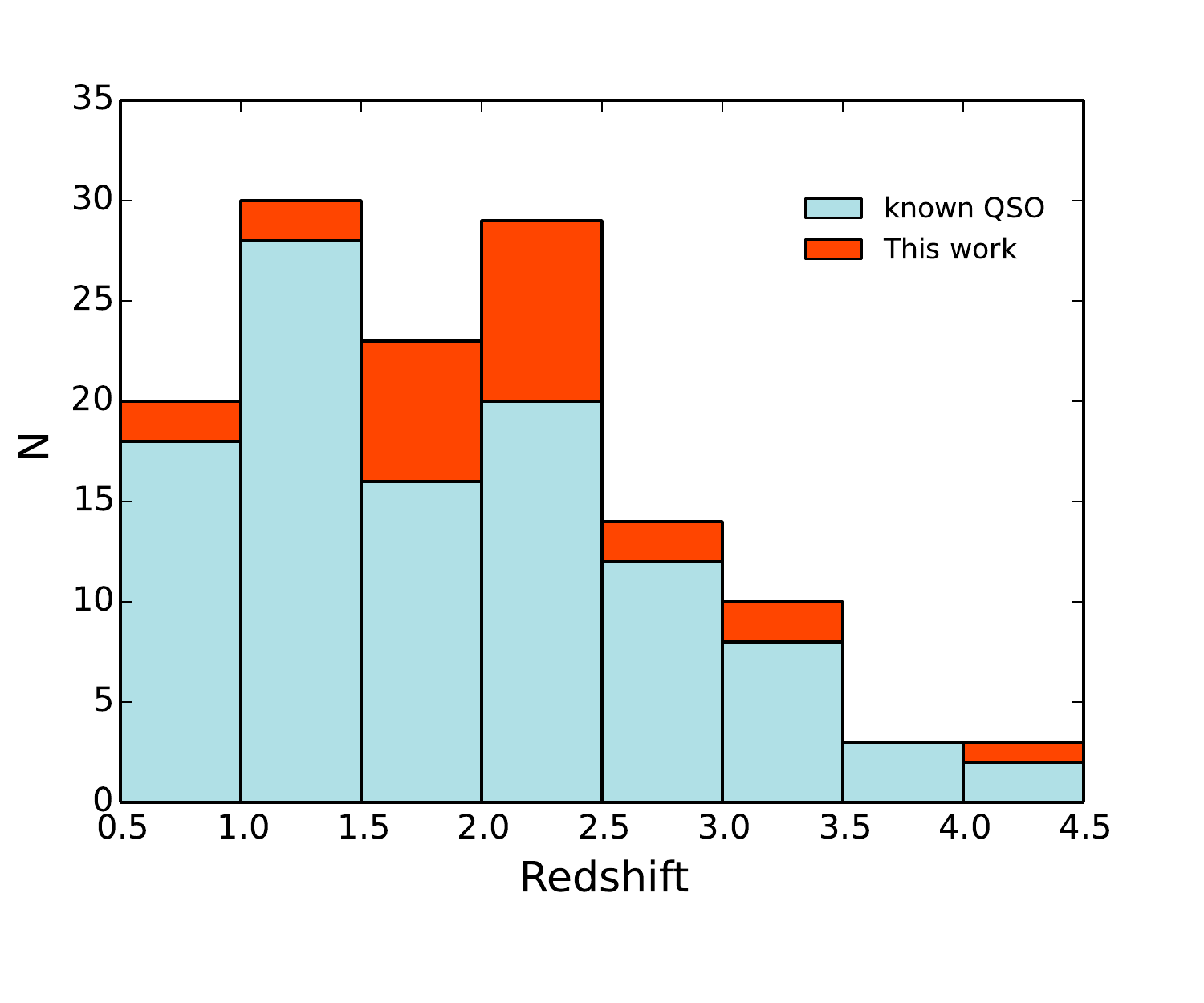}
\centering
\caption{Redshift distribution of newly discovered quasars and 110 previously known quasars. Our survey mainly contributes to the quasar distribution at $z \sim 2$.}
\label{fig4}
\end{figure}

We reduced all MMT/Hectospec spectra using the IDL pipeline HSRED v2.0\footnote{git://github.com/richardjcool/HSRed.git}.  
We used Stripe 82 $r$ band photometric data for absolute flux calibration. As a comparison, we re-observed three previously known SDSS quasars.  
From a comparison between MMT spectra and SDSS spectra, we believe that the slope, wavelength and flux calibration of our MMT spectra are reasonable. We measure the redshifts by visually matching the observed spectrum to quasar template using an eye-recognition assistant for quasar spectra software ASERA \citep{yuan13}. This matching is based on broad emission lines of $\rm Ly \beta$, $\rm Ly \alpha$, N\,{\sc v}, O\,{\sc i}/Si\,{\sc ii}, Si\,{\sc iv}, C\,{\sc iv}, C\,{\sc iii} and Mg\,{\sc ii}. The typical uncertainty of our redshift measurement is around 0.03. We finally obtain 25 new quasars with 0.6 $< z <$4.3. The MMT/Hectospec spectra of these new quasars are shown in Figure 3. Other candidates without broad emission lines can not be identified as quasar. The signal-to-noise ratio of those spectra are also not high enough for specific stellar types. Typically, the main contaminations for $z \sim 2-3$ quasars are A and F stars. Since we did not strictly limit the magnitude errors, large photometric errors will broaden the stellar locus and thus the contamination rate will be high. In addition, we relaxed star/galaxy separation to cover more quasars, so compact galaxy will also be a factor.

\begin{figure}%f5
\centering
\includegraphics[width=0.52\textwidth]{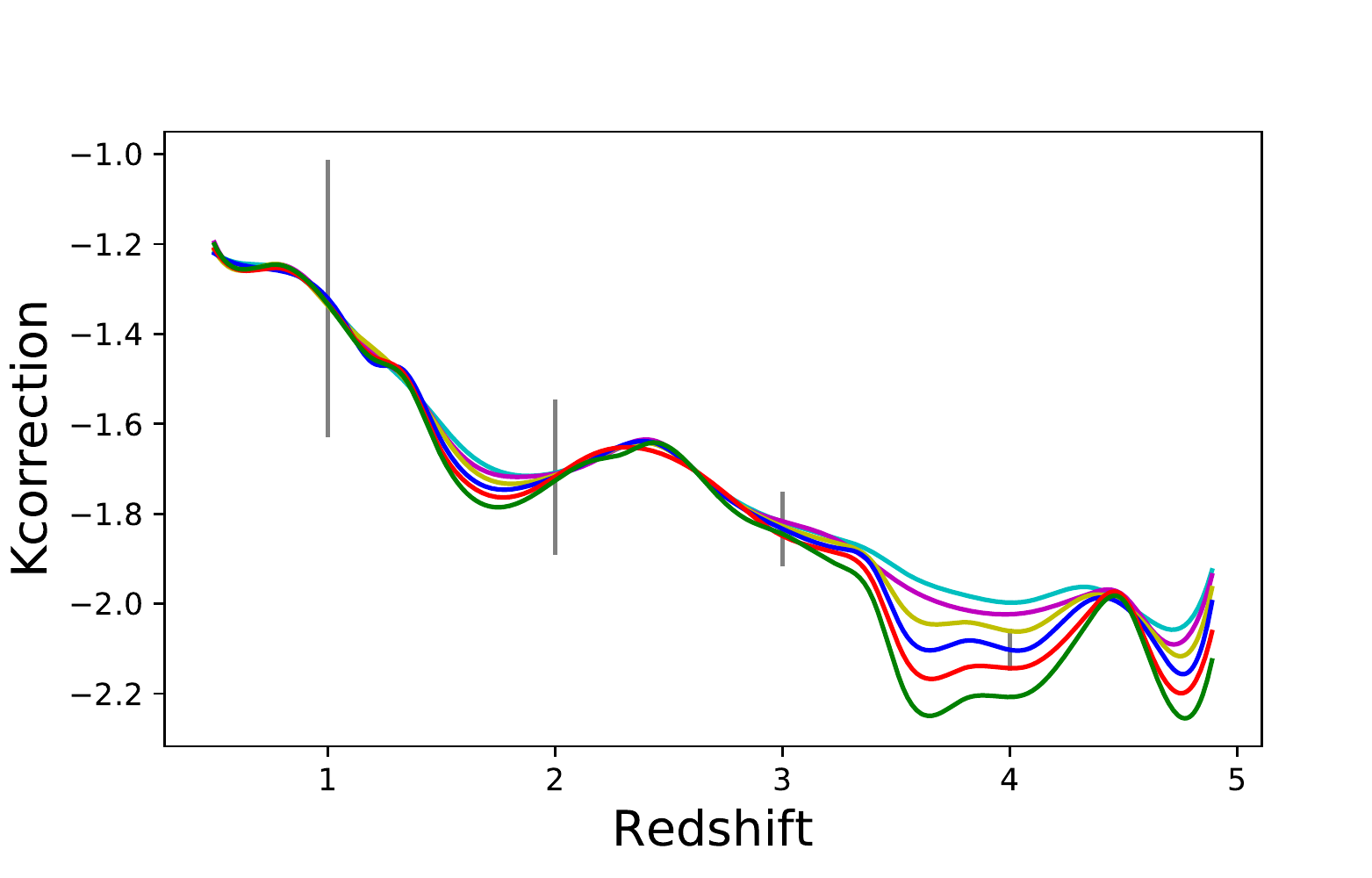}
\caption{k correction used to convert apparent $i$ band magnitude to absolute magnitude $M_{1450}$. We plot them in different $i$ band magnitude bins from $<i>$ = 17.5 (cyan) to 22.5 (green) in the step of $\Delta$m =1. The magnitude-dependent k correction is caused by the luminosity-dependent equivalent width of emission line. The gray vertical lines represent scatter of k correction at redshift $z = 1, 2, 3, 4$, generated by calculating standard deviation of k corrections in simulated spectra.}
\label{fig5}
\end{figure}

\subsection{A uniformly selected quasar sample}

Combining our new discoveries and previously known quasars, we construct a uniform sample of quasars at $0.5 < z < 4. 5$ within this 1 square degree field. We restrict the area to: $334.50^\circ \le RA \le 335.47^\circ$ and $-0.025^\circ \le Decl. \le 1.023^\circ$, a 1.02 square degree field. In this region, there are 114 previously known quasars in total in the redshift range of $0.5 < z < 4.5$, mainly from SDSS DR7, DR9 \&DR10 and VVDS spectroscopy \citep{schneider10, paris12, paris14, paris17, garilli08, lefevre13}. Four quasars were not detected by deep $Y$-band, Stripe 82 or UDXS photometry. The two missed by $Y$-band image located near the edge of image or in a masked region. In Figure 4, we plot the redshift distribution of all quasars in this field detected by Stripe 82, $Y$-band and UDXS photometry, including 110 known quasars and our new discoveries. As shown, our work has added a significant number of quasars at $z \sim 2$, which is one of the goals of our survey. Among those 110 known quasars, 107 quasars are brighter than our magnitude limit. All of these 107 quasars are selected by our color-color selection criteria. It also confirms the high completeness of our color-color selection which can cover a wide redshift range from 0.5 to 4, especially over the highly contaminated redshift range at $2 < z <3$. Comparing with BOSS \citep{ross12, ross13}, our color-color criteria yield higher completeness by involving only $g, i$ and $z$ bands in optical and adding NIR colors. 

There are 10 previously known quasars rejected by point/extended separator, 12 quasars rejected by $\chi^{2}$ limit, and another one quasar rejected by both. Our relaxed limit on magnitude errors will result in a lower successful rate but higher completeness of color-color selection. While objects with large photometric errors may be rejected by $\chi^{2}$ estimator since larger photometric error will lead to a larger $\chi^{2}$ value. The completeness of selection pipeline will be quantified in next Section. Our final uniformly selected quasar sample includes 109 quasars at $0.5 < z < 4.5$. By counting all selected known quasars and our new discoveries, we can estimate a contamination rate of our selection. There were 365 objects selected by our selection in this field, including 84 previously known quasars and our 281 candidates. Through the Hectospec observation, we observed 143 candidates and obtained 25 new quasars among 281 candidates. After applying the incompleteness correction of spectroscopy (See Sec. 4), we estimate that the number of expected quasars selected by our selection in this field is 130 in total. Therefore, we obtain a 64.4\% (1 $-$ 130/365) contamination rate of our selection method.

We calculate the absolute AB magnitude at the rest-frame 1450\AA\, $M_{1450}$ of all 109 quasars using k correction determined by using a sample of simulated quasar spectra. The simulated quasar sample is built based on quasar template with scatters of continuum slope and equivalent width of emission lines. This sample is what we use for the estimation of selection function in Section 4.1, where we will describe in more details. The k correction shows dependence on luminosity at some redshifts due to the luminosity-dependent equivalent width of emission line, e.g. the Baldwin effect \citep{baldwin77}. Since we use $i$ band apparent magnitude to estimate $M_{1450}$, we generate an $i$ band magnitude-dependent k correction. Using simulated quasars, we produce k-correction curves from $i$ band apparent magnitude to $M_{1450}$ at the redshift range of $0.5 < z < 4.5$, by calculating the mean value of k correction at each redshift and $i$ band magnitude bins ($\Delta z = 0.01$, $\Delta i = 0.01$), as shown in Figure 5. In Figure 5, we can see that the largest difference in k correction between different magnitude bins is at the redshift range of $3.5 < z < 4.4$, when C\,{\sc iv} line moves into $i$ band. At $z > 4.6$, the $i$ band magnitude is significantly affected by Ly$\alpha$ emission line, although Ly$\alpha$ somewhat shows weaker Baldwin effect than C\,{\sc iv} \citep{peterson97}. The difference shown at $z \sim 1.6$ represents the effect from Mg\,{\sc ii}. We list redshifts, $M_{1450}$ and photometry information of 109 quasars in Table 1. All optical data have been corrected for Galactic extinction.

\section{Selection Function}
%\subsection{Completeness}
In this section, we will describe the completeness of our selection pipeline, including photometry detection, point/extend source separation, color-color cuts, $\chi^{2}$ limit, and spectroscopy. As discussed above, there are 4 of 114 quasars missed by deep $Y$-band, Stripe 82 or UDXS photometry. So we directly use this fraction as the detection incompleteness of the Stripe 82-Y-UDXS photometric data set we used, by assuming that the candidate sample has missed the same fraction of quasars to known quasars. 

\begin{figure}%f6
\centering
\includegraphics[width=0.5\textwidth]{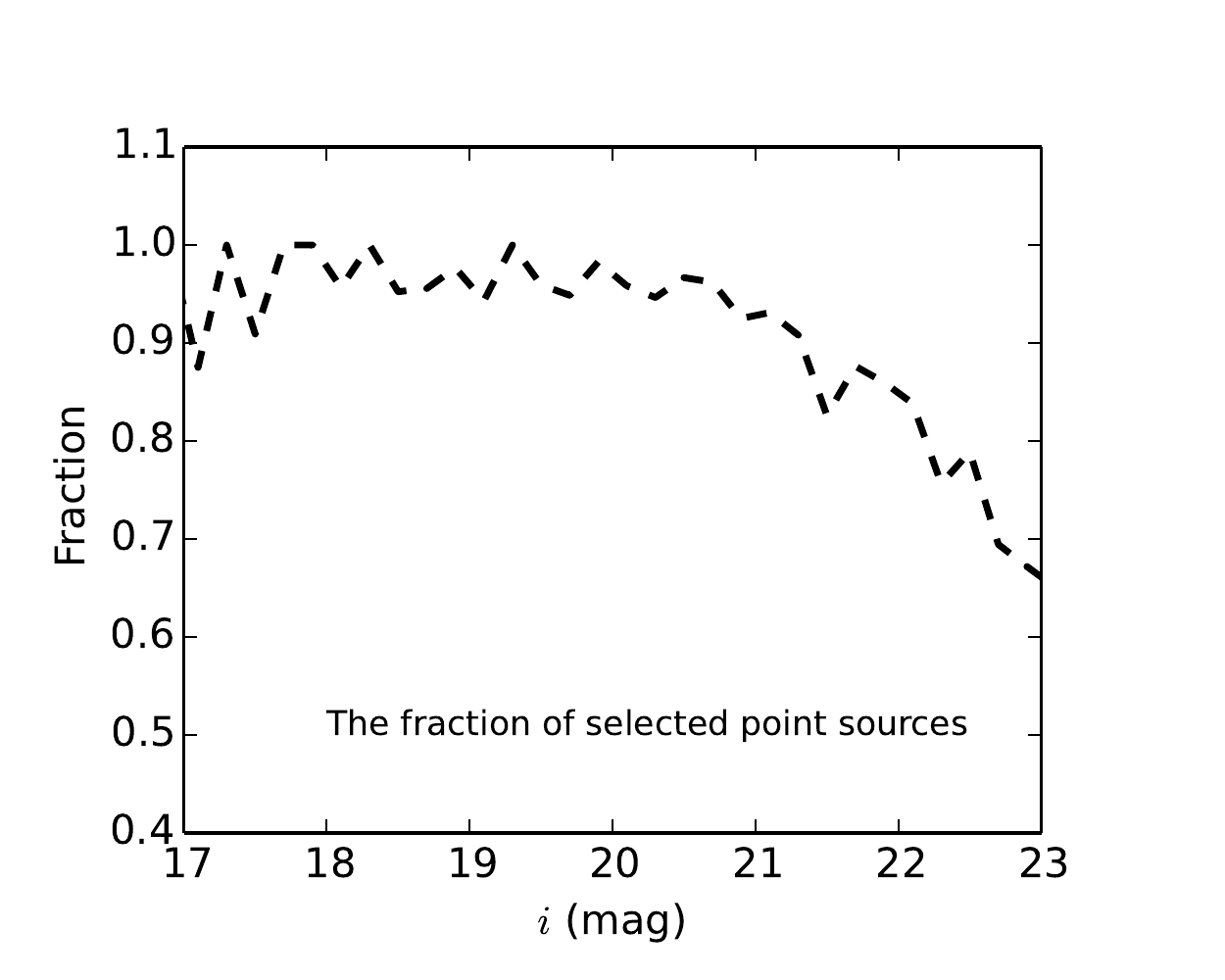}
\caption{The completeness of point/extend source separation from $i$ band magnitude 17 to 23 mag. It is a function of $i$ band magnitude.}
\end{figure}

The completeness of point/extended source separation is expected to be a function of brightness. We determine the completeness by using a sample of HST imaged point sources. These point sources are required to locate within Stripe 82 covered areas which have similar observing condition and image quality to our one square degree field. Based on Figure 4, 5 and 6 in \cite{annis14}, we choose the area in the range of $0^\circ < Decl. < 1^\circ$ in Stripe 82 region, in which $Decl.$ range photometry data have similar seeing condition, number of repeat observations, and zero point with our one square degree field. We select all HST detected point sources within this area from Hubble Source Catalog \citep{whitmore16} in HST Hubble Legacy Archive Release \footnote{http://archive.stsci.edu/hst/hsc/}. We then cross match (1") them with Stripe 82 catalog to get the photometry data. Using the final point source sample, including $\sim$ 2600 HST point sources, we calculate the fraction of how many point sources can be successfully classified as point source by our separation. It is a function of $i$ band magnitude, as shown in Figure 6. The completeness is almost higher than 90\% at $i$ band magnitude brighter than 21 and drops to 75\% to the magnitude limit ($i$ = 22.5) of our survey. The average completeness within the magnitude limit is 92\%, which is consistent with the selected fraction of previously known quasars (96/107) as discussed above.

To estimate the completeness of our color-color selection criteria, we generate a sample of simulated quasars following the procedure in \cite{fan99a} and \cite{mcgreer13}. \cite{fan99a} described the procedure to generate simulated quasar spectra using an empirical model for quasar spectral properties at UV and optical wavelength. Based on simulated spectra, we can measure the simulated colors of each spectrum by integrating spectrum through bandpasses used by survey. Simulated colors can be used to define selection criteria and estimate the completeness of color cuts. \cite{mcgreer13} updated the spectral model of \cite{fan99a} and applied it to higher redshift, assuming that the quasar spectral energy distributions do not evolve with redshift \citep{kuhn01, yip04, jiang06}. The quasar spectrum from \citep{mcgreer13} is modeled as a power law continuum with a break at 1100\AA. They used normal distributions to describe the continuum slopes. The distribution of the blue side is $\mu(\alpha)$ = -1.7 and $\sigma(\alpha)$ =0.3 \citep{telfer02}; the distribution of the red slope is $\mu(\alpha)$ = -0.5 and $\sigma(\alpha)$ = 0.3. Emission lines have been added to continuum using Gaussian profiles. The Gaussian parameters are also drawn from normal distributions which are generated from fitting composite spectra of quasars from the BOSS survey in different luminosity bins. We added breaks at 5700$\rm \AA$ and 9730$\rm \AA$ for redder wavelength coverage (in $J, K$ bands), following the similar procedures in \cite{yang16}. The slope ($\alpha_{\nu}$) from 5700$\rm \AA$ to 9730$\rm \AA$ follows a Gaussian distribution of $\mu$($\alpha$) = $-$0.48 and $\sigma$($\alpha$) = 0.3; the redder range continuum has a slope with the distribution of $\mu$($\alpha$) = $-$1.74 and $\sigma$($\alpha$) = 0.3 \citep{glikman06}. The parameters of emission lines are derived from the composite quasar spectra \citep{glikman06}.

\begin{figure}%f7
\centering
\includegraphics[width=0.55\textwidth]{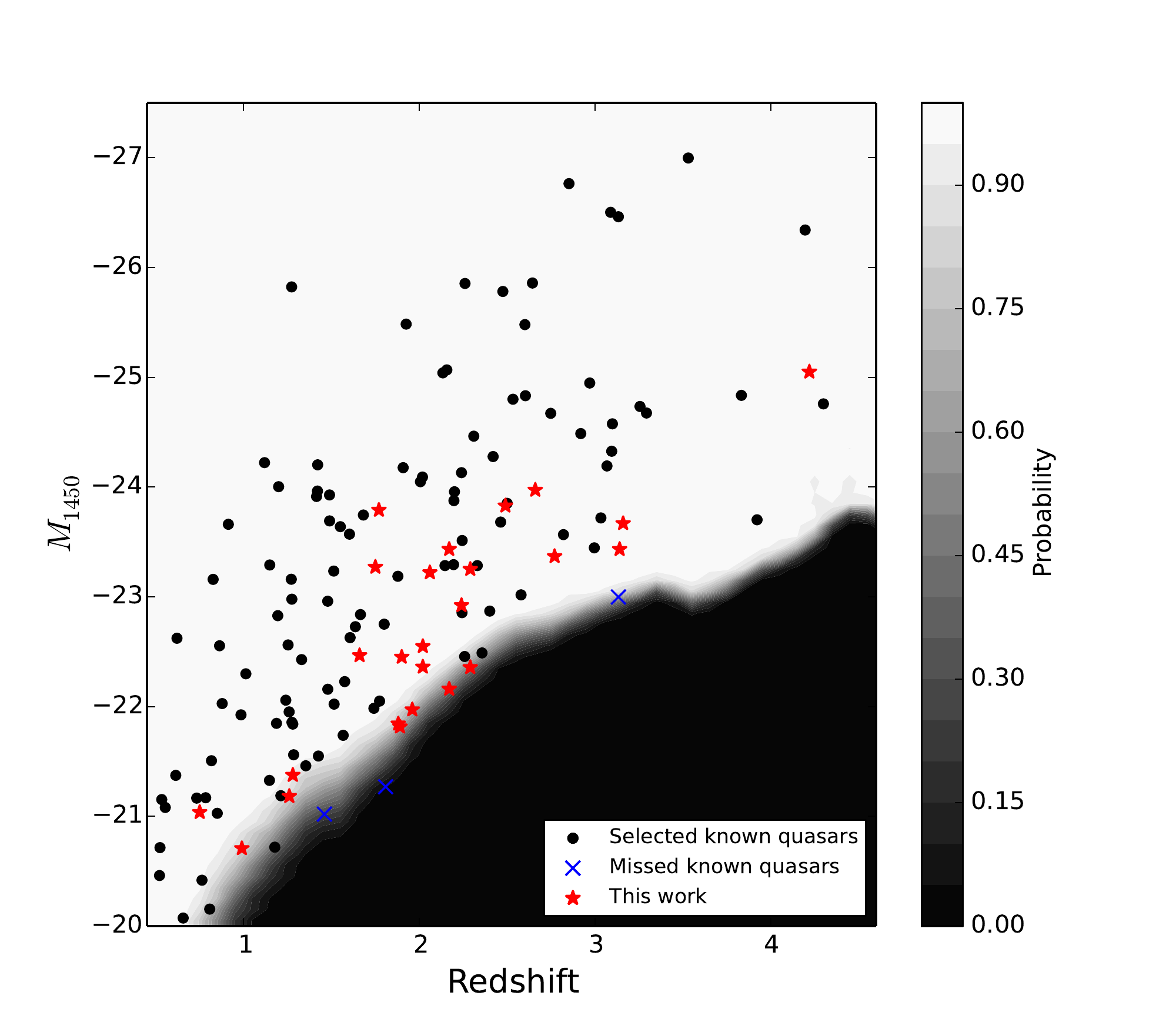}
\caption{Selection function of color-color selection criteria. All 110 Stripe 82-Y-UDXS detected known quasars are denoted by black filled circles and blue crosses, for selected and missed known quasars, respectively. Red stars represent our new discoveries. Our selections show a very high completeness at the full redshift range from $z = 0.5$ to 4.5. The completeness drops quickly at the bottom-right region, which is only caused by the magnitude limit ($i <$ 22.5) of our survey. All of three missed quasars are rejected by this magnitude limit. At the bright end, there should be a small region out of the magnitude range ($17 < i < 23.1$) of simulation sample (See fig.8 also). But, for plot, here we also set the completeness value in this region to 1.0 since based on the selection function in nearby area our color selections can be expected to be continuous and keep the high completeness.}

\label{fig7}
\end{figure}

The intergalactic medium absorption model used for our simulation is the same as \cite{mcgreer13}, which extend the Ly${\alpha}$ forest model based on the work of \cite{WP11} to higher redshift by using the observed number densities of high column density systems \citep{SC10}. Compared to \cite{mcgreer13}, we also made minor modifications for Fe emission. We use the template from \cite{vestergaard01} for wavelengths shorter than 2200\AA. For 2200-3500\AA, we use the template from \cite{tsuzuki06} which separates the FeII emission from the MgII $\lambda$2798 line. A template from \cite{boroson92} covering 3500-7500$\rm \AA$ is also added. The photometric datasets we used in optical and Y bands are from different surveys, and thus have different depths, with that were used by \cite{mcgreer13}. Therefore, we need to simulate photometric uncertainties of Stripe 82 $u, g, r, i, z$ photometry and our Y-band data. We use a sample of Stripe 82 classified point sources to fit the magnitude - error relations in $u, g, r, i, z$ and Y bands. Using the simulation model and new magnitude - error relations, we generate photometric data of 1,085,800 simulated quasar spectra, evenly distributed in the ($i$, $z$) space. We construct a grid of quasars in the redshift range of $0.5 < z < 4.95$ and the magnitude range of $17 < i < 23.1$ to match the magnitude and redshift ranges of our quasar sample. There are $\sim$ 200 quasars in each ($i$, $z$) bin with $\Delta m$= 0.1 and $\Delta z$= 0.05.

\begin{figure}%f8
\centering
\includegraphics[width=0.55\textwidth]{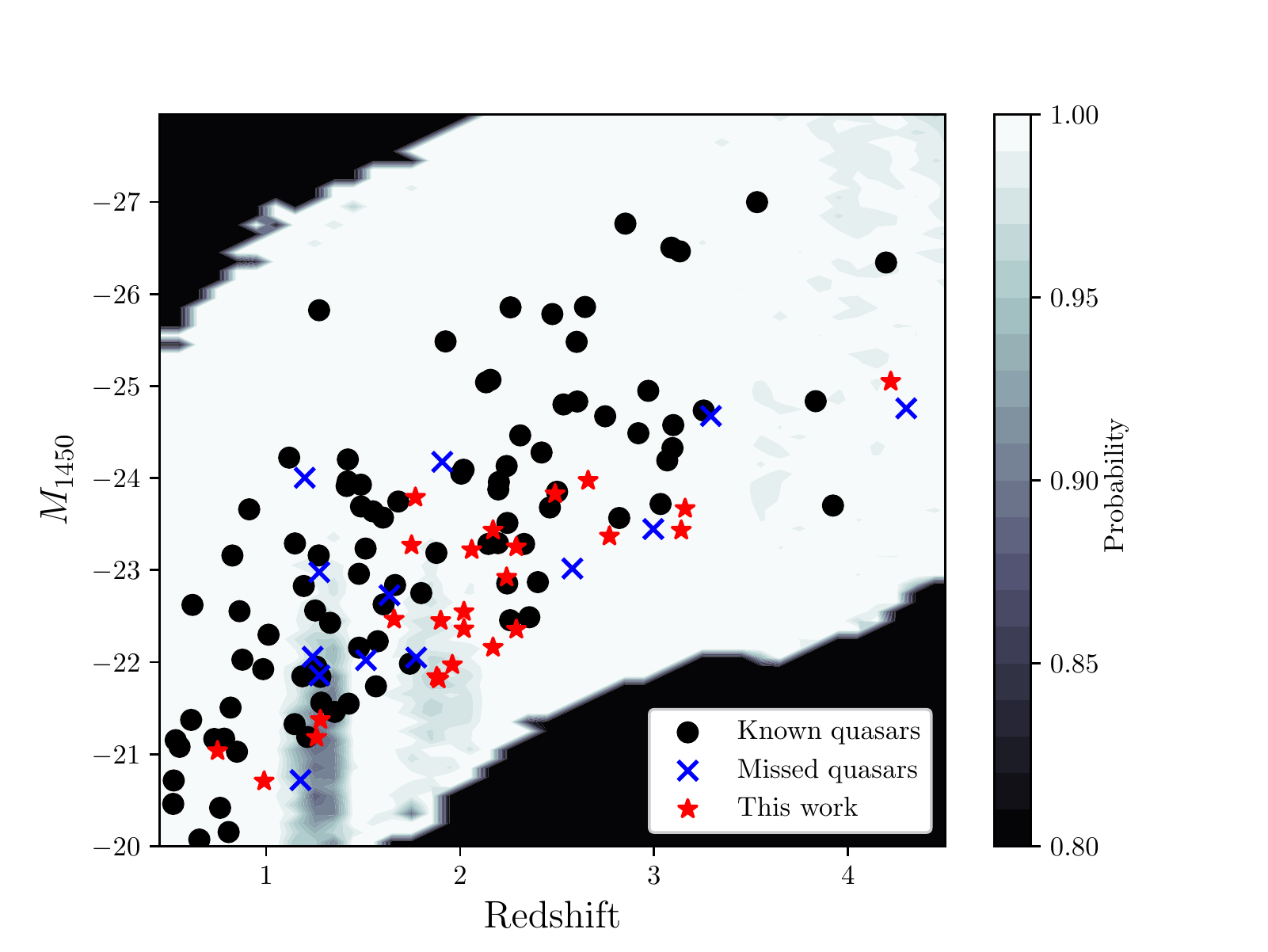}
\caption{The completeness of $\chi^{2} < 15$ limit in the magnitude range of $17 < i < 23.1$. The two dark regions are out of the magnitude range. Red stars represent our new discoveries. All previously known quasars are denoted by black filled circles and blue crosses, for selected and missed quasars, respectively. Considering more than 99.85\% points show a completeness higher than 83.3\%, here we scale the color bar to 0.8-1.0. }
\end{figure}

Based on the simulated quasar sample, we calculate the fraction of simulated quasars selected by our selection criteria as the selection probability, shown in Figure 7. We calculate the probability in each ($M_{1450}$, $z$) bin in the range of $-28 < M_{1450} < -20$ and $0.5 < z < 4.95$ with $\Delta M$= 0.1 and $\Delta z$= 0.1. As shown in Figure 7, our color-color selection criteria is highly complete at the magnitude range brighter than our magnitude limit ($i < 22.5$). It is consistent with the selection fraction of known quasars: all of three missed known quasars are only rejected by $i$ band magnitude limit. Using this simulated quasar sample, we also calculate the incompleteness of $\chi^{2}$ fitting. We calculate $\chi^{2}$ value of each simulated quasar using the same method that we used for candidates selection. We generate the selection probability of $\chi^{2}$ limit in ($M_{1450}$, z) space using the same bins as discussed above. The completeness of $\chi^{2}$ estimator is plotted in Figure 8. When we constructed the color-z relation for $\chi^{2}$, to make the color-z accurate we reject quasars out of 3 $\sigma$ to the mean values of colors. The simulation method also does not include unusual weak line quasar and broad absorption line quasar. Therefore, quasars with extreme colors will not be covered by both simulation quasar sample and $\chi^{2}$ estimator. Thus we will get a higher completeness of $\chi^{2}$ limit than truth by using the simulation quasar sample. The difference will be smaller than the uncertainties of QLF measurement. We finally estimate the spectroscopy incompleteness by assuming the same fraction of quasars in observed and unobserved candidates sub-samples. Since when we did fiber allocation, we chose a configuration that could target more rank 1 candidates, here we estimate the spectroscopy incompleteness of each rank respectively. The spectroscopy completeness of quasars in rank 1 is 67\%, while it is 33\% and 30\% for rank 2 and rank 3 quasars, respectively.

\section{QLF in one square degree field}
To draw the distribution of quasars at different redshifts and magnitudes in this one square degree field, we calculate the QLF using this quasar sample including 109 quasars. We calculate the binned luminosity function by using the \cite{page-carrera00} implementation of the traditional 1/$\rm V_{max}$ method \citep{schmidt68, avni-bahcall80} for flux limit correction. We divide our sample into 8 redshift bins at $0.5 < z < 4.5$ ($\Delta z=0.5$) and 7 $M_{1450}$ magnitude bins at $-27 < M_{1450} < -20$ ($\Delta M =1 $). Since our quasar sample in a small area only includes 109 quasars but covers a wide redshift and magnitude range, if we use smaller magnitude bin, there will be one or two quasars in each bin, leading to large uncertainties of binned QLF. We prefer to focus on the redshift evolution and thus choose a large magnitude interval with $\Delta M =1$ mag. We correct the incompleteness of our selection as discussed above. The binned data are listed in Table 2 and plotted in Figure 9. Considering the small quasar sample size in each $M_{1450}-z$ bin, we use mean value of redshift and $M_{1450}$.

\begin{figure*}%f9
\includegraphics[width=0.9\textwidth]{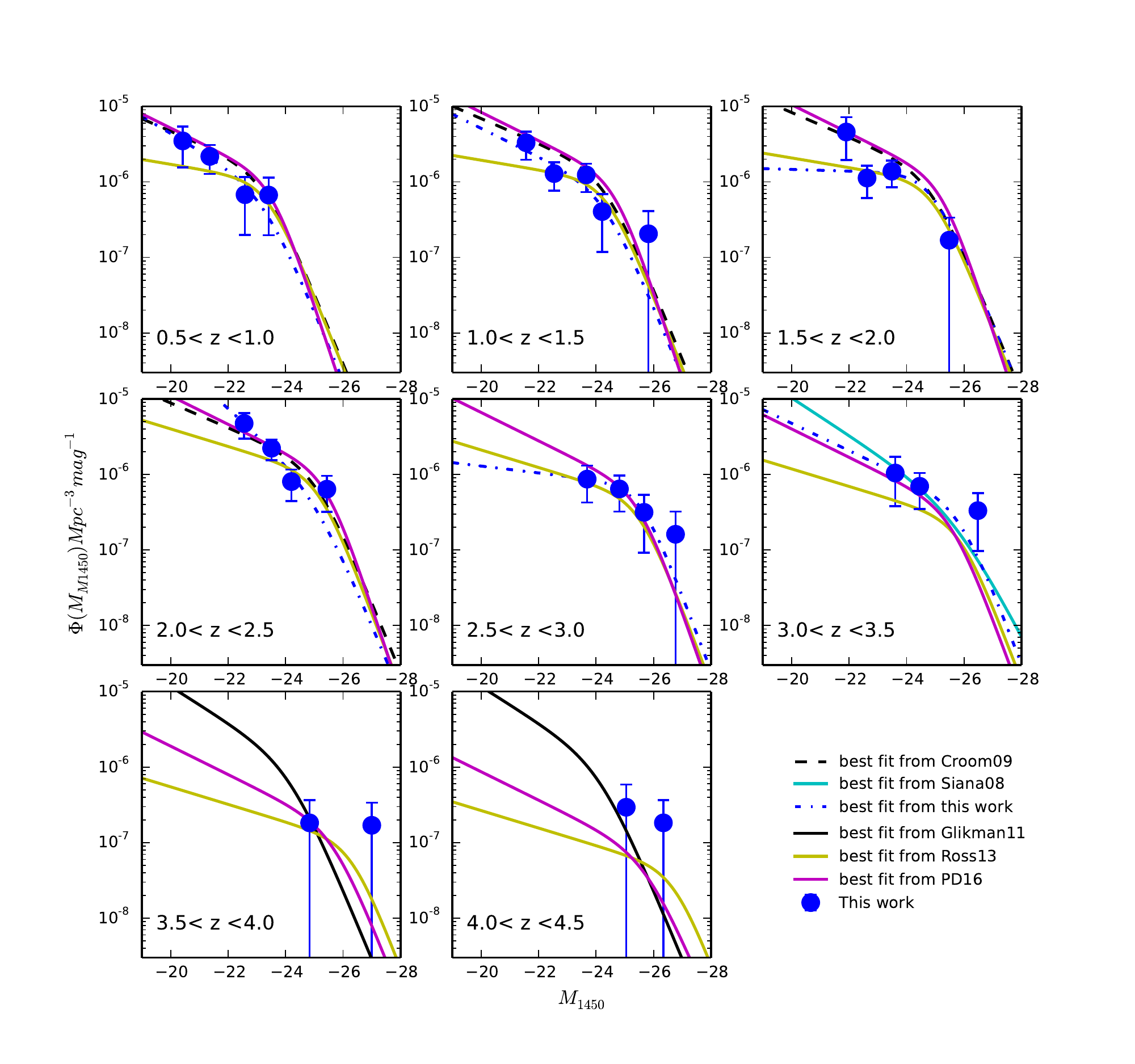}
\centering
\caption{The binned (blue filled circle with error bars) and best-fits (blue dashed line) QLFs form our quasar sample. As a comparison, we plot QLFs from previous works at different redshifts. We compare our QLFs with the best-fit QLFs from \cite{croom09} at bins with $z < 2.5$ (black dashed lines), \cite{ross13} (yellow solid lines) with a PLE model at bins with $z < 2 $ and a LEDE model at $z > 2$ (using parameters from Stripe 82 sample), \cite{Palanque-Delabrouille16} (purple solid lines) in the full redshift range, the $z \sim 3$ QLF from \cite{siana08} (cyan lines) at redshift bin $3 < z < 3.5$ and the $z \sim 4$ QLF from \cite{glikman11} (black solid lines) at bins with $3.5 < z < 4.5$. We have converted the magnitudes used in their QLFs into $M_{1450}$ and the cosmology into our adopted cosmology.}
\end{figure*}

Since our quasar sample cover a wide redshift range, we also measure the parametric QLF to discuss the evolution model of QLF with redshift. We model the parametric QLF following the double power law form \citep{boyle00}:
 \begin{equation}
	\Phi(M,z) = \frac{\Phi^*(z)}
	     {10^{0.4(\alpha+1)(M-M^*)} + 10^{0.4(\beta+1)(M-M^*)}} ~
\end{equation}   
where $\alpha$ and $\beta$ are the faint end and the bright end slopes; $M^{*}$ is the break magnitude and $\Phi^*(z)$ is the normalization. We use the $\chi^{2}$ fitting to fit binned QLF data at each redshift bin with the double power law formula. Our sample only covers the faint end of each redshift bin, so we fix the bright end slope $\beta$ and break magnitude $M^{*}_{1450}$. Previous works have show strong evidence of a pure luminosity evolution (PLE) model of QLF at $z \lesssim 2.2$ and a luminosity evolution and density evolution (LEDE) model of QLF at $z \gtrsim 2.2$ \cite{croom09, ross13, Palanque-Delabrouille13, Palanque-Delabrouille16}. \cite{croom09} used a sample of quasars from 2dF-SDSS LRG and QSO (2SLAQ) survey to derive the QLF at $0.4 < z < 2.6$. The PLE model used by \cite{croom09} shows the redshift dependence through the evolved break magnitude, described by
\begin{equation}
M^{*}(z)  = M^{*}(0) - 2.5(k_{1}z + k_{2}z^{2})~
\end{equation}   
For bins at $z < 2.5$, we choose the same formula of the PLE model from \cite{croom09}, and thus fix the bright end slope $\beta$ and break magnitude $M^{*}_{1450}$ to their result. 

At $z > 2.5$, we choose the LEDE model from \cite{ross13}. \cite{ross13} used the BOSS color selected DR9 quasar sample and BOSS Stripe 82 variability selected quasar sample to measure the QLF at 2.2 $< z <$ 3.5, and conclude that the QLF can be described well by an LEDE model, which is in a log-linear formula

\begin{eqnarray}
  \log[\Phi^{*}(z)]  & = & \log[\Phi^{*}(z=2.2)] + c_{1}(z-2.2)~,
  \label{eq:LEDE_Phistar} \\
            M_{i,2}^{*}(z) & = &M_{i,2}^{*}(z=2.2) + c_{2}(z-2.2)~
  \label{eq:LEDE_Mstar}
\end{eqnarray} 
We also fix $\beta$ and $M^{*}_{1450}$ to the result given by \cite{ross13}. Because in our sample there are only few luminous quasars in each bin at the bright end, the bright end slopes will be highly uncertain due to small number statistics. Therefore, when we do the fitting, at each redshift bin, we reject binned QLF data in the magnitude bins that are brighter than the break magnitude. We do not do the fitting for bins at $z > 3.5$ since there are also only few quasars. The best-fits at different bins are summarized in Table 2.

In Figure 9, we plot our best-fit QLFs, comparing with binned QLF and QLFs at different redshifts from previous works \citep{croom09, glikman11, Palanque-Delabrouille16, ross13, siana08}. 
The result from \cite{croom09} focused on the redshift range of $z < 2.6$, so we plot their QLF at bins of $z < 2.5$. \cite{ross13} concentrated on the redshift range of $2.2 < z < 3.5$, using both 23300 color selected quasars and 5476 variability selected quasars at $2.2 < z < 3.5$, and supplemented it with a deeper dataset over a smaller area to probe lower redshift at $0.7 < z < 2.2$. So we plot their result with PLE model at $z < 2.2$ and LEDE model at $2.2 < z < 3.5$. We also plot their LEDE QLF at $z > 3.5$ as a comparison, since the LEDE model with log-linear manner has also been suggested by the studies of high redshift QLF \citep[e.g.][]{mcgreer13,yang16}. \cite{Palanque-Delabrouille16} used the variability selected quasar sample in Stripe 82 area from SDSS-IV/eBOSS to present a determination of QLF at $0.6 < z < 4$. Their data could be described well by both PLE and PLE ($z < 2.2$) + LEDE ($z > 2.2$) model. Here we plot the PLE+LEDE model from \cite{Palanque-Delabrouille16}. They adopted the same PLE formula to \cite{croom09} and a similar log-linear LEDE model with \cite{ross13} but required the QLF to be continuos at the pivot redshift ($z_{p} = 2.2$). As shown, our binned QLF and best-fits show agreement well with the PLE model used by \cite{croom09, Palanque-Delabrouille16} at $z < 2.5$.  At $2.5 < z < 3.5$, our QLF is following the LEDE model and is more consistent with \cite{Palanque-Delabrouille16}. We also plot the $z \sim 3.2$ QLF from the Spitzer Wide-area Infrared Extragalactic (SWIRE) survey \citep[][;SWIRE+SDSS]{siana08} and $z \sim 4$ QLF from NOAO Deep Wide-Field Survey (NDWFS) + Deep Lens Survey (DLS) \citep{glikman11} at bins with $3.0 < z < 3.5$ and $3.5 < z < 4.5$ respectively. At $z \sim 3$, the QLF from \cite{siana08} is included in the 1$\sigma$ region of our binned data. At $z \sim 4$, our data have uncertainties too large to constraint the QLF. We show the $log(\Phi^{*})$ and the faint end slope $\alpha$ of best-fits versus redshift in Figure 10. Although our result is based on a small sample, the parameters evolution still can show the trend that $log(\Phi^{*})$ evolved with increasing redshift by a PLE model at lower redshift and a LEDE model at higher redshift. The faint end slope does not obviously evolve.

\begin{figure}%f10
\centering
\includegraphics[width=0.5\textwidth]{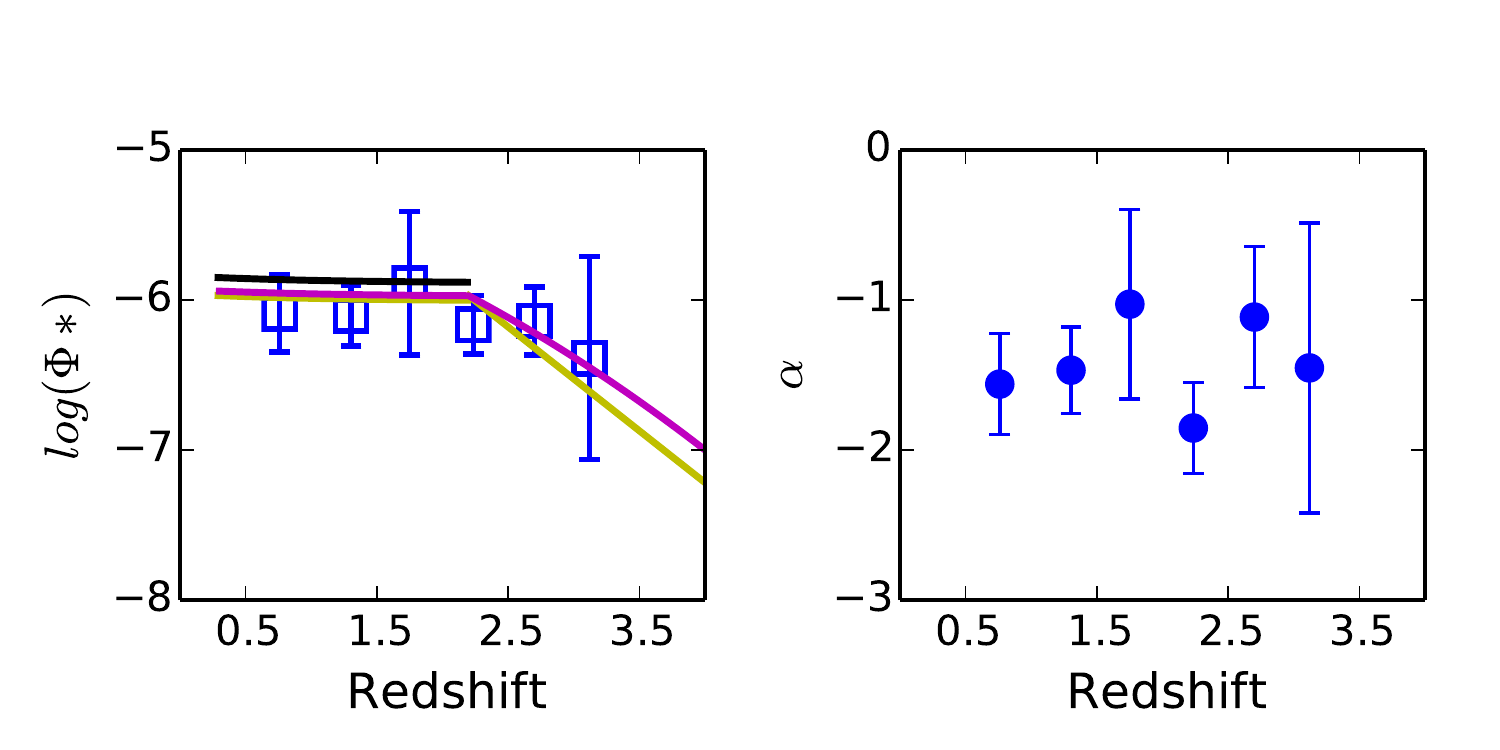}
\caption{Parameters evolution, $log(\Phi^{*})$ (left) and the faint end slope $\alpha$ (right). The yellow and black solid lines represent the PLE + LEDE evolution model from \cite{ross13} and PLE model given by \cite{croom09}. The PLE+LEDE model from \cite{Palanque-Delabrouille16} is also shown as comparison (purple line). }
\end{figure}

\section{Summary}
In this work,  we used deep CFHT Y-band image, deep optical data from SDSS Stripe 82 and NIR data from UKIDSS DXS DR9 to survey quasars in a one square degree field. We used the color-color selection criteria $Y-K$/$g-z$ and $J-K$/$i-Y$ \citep{wu-jia10} to select quasar candidates and discovered 25 new quasars at redshift range of $0.5 < z < 4.3$, which make obvious contribution to the quasars distribution of both $2 < z < 3$ and faint quasars in this field. By combining our new quasars and previously known quasars that meet our selection pipeline, we construct a quasar sample including 109 quasars in a 1.02 $deg^{2}$ field. We estimate the completeness of our selection pipeline using a sample of simulated quasars. It confirms that the $Y-K$/$g-z$ and $J-K$/$i-Y$ color-color cuts are highly complete at $z< 4.5$: all quasars within the magnitude limit can be selected by color-color criteria. We calculate the QLFs, both binned and parametric QLFs. The results show agreements with the PLE evolution model at $z < 2.5$ and the LEDE evolution at $z > 2.5$. 

This optical+NIR color selection, yielding a high completeness at the wide redshift range of $z < 4.5$, can be applied for large area quasar survey to provide large complete quasar sample, especially for $2 < z < 3$ quasars. This method has already been used by the Large Sky Area Multi-Object Fiber Spectroscopic Telescope (LAMOST) quasar survey \citep[][;X. Dong 2017 in prep]{ai16} in the entire SDSS-ULAS area. More than 2500 new quasars have been discovered by using this selection. The new optical and NIR surveys with wide sky coverage, e.g. Pan-STARRS 1 \citep[PS1;][]{chambers16}, the VLT Survey Telescope ATLAS survey \citep{shanks15} and the VISTA Hemisphere Survey \citep[VHS;][]{mcmahon13}, will offer prefect optical+NIR photometric dataset for a complete $z < 4.5$ quasar survey as well.

\vspace{7pt}
\LongTables
\begin{deluxetable*}{l c l c c c c c c c}
\centering
\small
 \tablecaption{Quasar sample in one square degree field}
% \tablewidth{0pt}
 \tablehead
{ \colhead{Name} &
  \colhead{Redshift} &
  \colhead{$Ref_{z}$} &
  \colhead{$M_{1450}$} &
  \colhead{$g$} &
  \colhead{$i$} &
  \colhead{$z$} &
  \colhead{$Y$} &
  \colhead{$J$} &
  \colhead{$K$}
 } 
\startdata
J221801.63+000041.18 & 1.665 & SDSS & $-$22.84 & 21.21$\pm$0.012 & 20.96$\pm$0.018 & 20.91$\pm$0.076 & 20.43$\pm$0.026 & 19.67$\pm$0.027 & 18.59$\pm$0.030\\ 
J221805.79+000912.38 & 1.878 & VVDS & $-$23.19 & 21.46$\pm$0.014 & 20.94$\pm$0.017 & 20.82$\pm$0.068 & 20.43$\pm$0.024 & 19.79$\pm$0.028 & 18.45$\pm$0.026\\ 
J221806.61+000535.08 & 2.310 & SDSS & $-$24.46 & 20.24$\pm$0.006 & 20.21$\pm$0.009 & 19.96$\pm$0.032 & 19.45$\pm$0.009 & 19.25$\pm$0.020 & 17.87$\pm$0.017\\ 
J221806.68+005223.73 & 1.273 & SDSS & $-$25.82 & 17.36$\pm$0.001 & 17.26$\pm$0.001 & 17.34$\pm$0.003 & 17.00$\pm$0.003 & 16.78$\pm$0.005 & 15.74$\pm$0.004\\ 
J221807.92+005229.82 & 3.095 & SDSS & $-$24.33 & 21.52$\pm$0.016 & 21.13$\pm$0.023 & 21.19$\pm$0.084 & 20.66$\pm$0.026 & 20.13$\pm$0.033 & 19.34$\pm$0.052\\ 
J221809.38+004133.48 & 1.660 & This work & $-$22.47 & 21.78$\pm$0.017 & 21.33$\pm$0.024 & 21.23$\pm$0.070 & 20.70$\pm$0.027 & 20.11$\pm$0.032 & 18.55$\pm$0.027\\ 
J221810.33+005017.19 & 1.606 & SDSS & $-$22.63 & 21.51$\pm$0.015 & 21.08$\pm$0.021 & 21.20$\pm$0.083 & 20.67$\pm$0.029 & 20.06$\pm$0.031 & 18.59$\pm$0.028\\ 
J221812.92+002628.39 & 1.259 & SDSS & $-$21.95 & 21.94$\pm$0.021 & 21.09$\pm$0.021 & 21.03$\pm$0.062 & 20.34$\pm$0.018 & 19.80$\pm$0.027 & 18.35$\pm$0.024\\ 
J221813.42+004854.03 & 2.357 & VVDS & $-$22.49 & 23.06$\pm$0.054 & 22.24$\pm$0.057 & 21.88$\pm$0.130 & 21.21$\pm$0.038 & 20.87$\pm$0.055 & 19.65$\pm$0.068\\ 
J221814.21+002049.66 & 1.513 & SDSS & $-$23.23 & 20.46$\pm$0.006 & 20.31$\pm$0.009 & 20.33$\pm$0.041 & 19.78$\pm$0.030 & 19.50$\pm$0.022 & 18.22$\pm$0.022\\ 
J221814.58+002736.82 & 2.240 & SDSS & $-$24.13 & 20.76$\pm$0.008 & 20.47$\pm$0.012 & 20.18$\pm$0.029 & 19.60$\pm$0.013 & 19.38$\pm$0.021 & 17.92$\pm$0.018\\ 
J221815.32+000117.65 & 2.533 & SDSS & $-$24.80 & 20.26$\pm$0.006 & 20.12$\pm$0.009 & 19.90$\pm$0.031 & 19.16$\pm$0.008 & 19.06$\pm$0.018 & 17.82$\pm$0.017\\ 
J221816.22+005848.36 & 1.330 & SDSS & $-$22.43 & 20.88$\pm$0.017 & 20.76$\pm$0.040 & 20.45$\pm$0.063 & 19.59$\pm$0.020 & 19.29$\pm$0.019 & 17.88$\pm$0.016\\ 
J221816.60+000701.30 & 1.742 & VVDS & $-$21.98 & 22.98$\pm$0.061 & 21.94$\pm$0.054 & 21.60$\pm$0.139 & 20.53$\pm$0.026 & 20.09$\pm$0.035 & 18.50$\pm$0.027\\ 
J221822.67+001715.32 & 0.524 & VVDS & $-$20.71 & 20.41$\pm$0.006 & 19.98$\pm$0.007 & 19.91$\pm$0.028 & 19.72$\pm$0.013 & 18.82$\pm$0.015 & 17.79$\pm$0.017\\ 
J221825.03+002426.56 & 0.913 & SDSS & $-$23.66 & 18.67$\pm$0.004 & 18.52$\pm$0.005 & 18.49$\pm$0.015 & 17.64$\pm$0.003 & 16.69$\pm$0.005 & 16.11$\pm$0.006\\ 
J221829.05+002024.14 & 1.479 & VVDS & $-$22.16 & 21.77$\pm$0.017 & 21.32$\pm$0.022 & 21.30$\pm$0.099 & 20.35$\pm$0.039 & 20.09$\pm$0.032 & 18.52$\pm$0.026\\ 
J221830.06+000005.07 & 3.256 & SDSS & $-$24.73 & 21.22$\pm$0.012 & 20.85$\pm$0.016 & 20.86$\pm$0.059 & 20.12$\pm$0.020 & 19.50$\pm$0.024 & 18.46$\pm$0.026\\ 
J221832.76+002424.94 & 1.890 & This work & $-$21.82 & 23.27$\pm$0.065 & 22.32$\pm$0.055 & 21.99$\pm$0.187 & 21.09$\pm$0.037 & 20.36$\pm$0.038 & 18.78$\pm$0.032\\ 
J221833.31+001835.06 & 1.147 & VVDS & $-$21.33 & 21.87$\pm$0.019 & 21.46$\pm$0.026 & 21.38$\pm$0.107 & 20.70$\pm$0.036 & 20.70$\pm$0.048 & 18.53$\pm$0.027\\ 
J221833.73+002709.46 & 1.253 & VVDS & $-$22.56 & 20.77$\pm$0.008 & 20.47$\pm$0.012 & 20.38$\pm$0.035 & 19.64$\pm$0.009 & 19.24$\pm$0.019 & 17.68$\pm$0.015\\ 
J221835.26+004839.16 & 2.290 & This work & $-$22.36 & 22.61$\pm$0.036 & 22.29$\pm$0.061 & 22.05$\pm$0.152 & 21.32$\pm$0.057 & 21.15$\pm$0.068 & 19.57$\pm$0.056\\ 
J221835.89+000342.44 & 3.068 & SDSS & $-$24.19 & 21.35$\pm$0.015 & 21.24$\pm$0.023 & 21.15$\pm$0.098 & 20.44$\pm$0.025 & 20.08$\pm$0.036 & 18.80$\pm$0.036\\ 
J221839.12+004945.40 & 1.280 & SDSS & $-$21.84 & 21.58$\pm$0.015 & 21.25$\pm$0.024 & 21.35$\pm$0.081 & 20.60$\pm$0.026 & 20.27$\pm$0.036 & 18.92$\pm$0.033\\ 
J221840.14+010038.52 & 2.970 & SDSS & $-$24.95 & 20.56$\pm$0.008 & 20.40$\pm$0.013 & 20.45$\pm$0.048 & 19.79$\pm$0.024 & 19.59$\pm$0.023 & 18.57$\pm$0.027\\ 
J221841.02+005311.01 & 3.140 & This work & $-$23.43 & 22.59$\pm$0.040 & 22.05$\pm$0.054 & 22.00$\pm$0.178 & 21.04$\pm$0.035 & 20.46$\pm$0.040 & 19.42$\pm$0.049\\ 
J221845.94+000953.06 & 2.258 & SDSS & $-$22.46 & 22.22$\pm$0.030 & 22.15$\pm$0.053 & 22.08$\pm$0.227 & 21.27$\pm$0.066 & 21.06$\pm$0.074 & 19.37$\pm$0.056\\ 
J221852.62+005740.39 & 2.401 & SDSS & $-$22.87 & 21.98$\pm$0.025 & 21.91$\pm$0.051 & 21.56$\pm$0.126 & 21.21$\pm$0.042 & 20.74$\pm$0.050 & 19.23$\pm$0.045\\ 
J221854.26-000058.75 & 3.099 & SDSS & $-$24.58 & 21.14$\pm$0.012 & 20.88$\pm$0.016 & 20.69$\pm$0.050 & 20.31$\pm$0.033 & 19.50$\pm$0.024 & 18.37$\pm$0.025\\ 
J221854.37+002656.97 & 0.733 & SDSS & $-$21.17 & 20.61$\pm$0.007 & 20.41$\pm$0.012 & 20.08$\pm$0.027 & 19.63$\pm$0.011 & 19.10$\pm$0.018 & 17.38$\pm$0.013\\ 
J221858.90+000712.21 & 3.034 & VVDS & $-$23.72 & 22.42$\pm$0.036 & 21.68$\pm$0.035 & 21.51$\pm$0.133 & 20.66$\pm$0.027 & 20.51$\pm$0.049 & 19.37$\pm$0.056\\ 
J221859.21+003927.50 & 1.489 & SDSS & $-$23.69 & 20.15$\pm$0.005 & 19.81$\pm$0.006 & 19.88$\pm$0.021 & 19.08$\pm$0.011 & 18.99$\pm$0.017 & 17.75$\pm$0.016\\ 
J221900.65+000812.26 & 1.415 & SDSS & $-$23.91 & 19.67$\pm$0.004 & 19.45$\pm$0.005 & 19.54$\pm$0.022 & 19.24$\pm$0.010 & 18.67$\pm$0.015 & 17.57$\pm$0.015\\ 
J221901.41+000610.72 & 0.863 & SDSS & $-$22.55 & 19.73$\pm$0.004 & 19.47$\pm$0.005 & 19.32$\pm$0.019 & 18.72$\pm$0.006 & 17.77$\pm$0.009 & 16.85$\pm$0.010\\ 
J221901.87+000025.92 & 1.149 & SDSS & $-$23.29 & 19.75$\pm$0.004 & 19.51$\pm$0.005 & 19.51$\pm$0.023 & 19.09$\pm$0.009 & 18.66$\pm$0.015 & 17.36$\pm$0.013\\ 
J221907.94+004023.48 & 0.614 & SDSS & $-$21.37 & 20.32$\pm$0.005 & 19.74$\pm$0.006 & 19.71$\pm$0.018 & 18.59$\pm$0.012 & 18.55$\pm$0.014 & 17.01$\pm$0.010\\ 
J221910.54+005606.50 & 1.195 & SDSS & $-$22.83 & 20.59$\pm$0.007 & 20.07$\pm$0.009 & 20.16$\pm$0.033 & 19.56$\pm$0.032 & 19.18$\pm$0.018 & 18.00$\pm$0.016\\ 
J221912.11+00361300 & 0.826 & SDSS & $-$23.16 & 18.72$\pm$0.002 & 18.75$\pm$0.003 & 18.65$\pm$0.008 & 18.79$\pm$0.018 & 17.49$\pm$0.008 & 16.48$\pm$0.008\\ 
J221912.60+000411.53 & 1.603 & SDSS & $-$23.57 & 20.49$\pm$0.007 & 20.13$\pm$0.009 & 20.29$\pm$0.044 & 19.94$\pm$0.055 & 19.27$\pm$0.021 & 17.95$\pm$0.020\\ 
J221915.76+004232.7 & 2.170 & This work & $-$23.43 & 21.18$\pm$0.010 & 21.08$\pm$0.019 & 20.79$\pm$0.046 & 20.18$\pm$0.019 & 20.18$\pm$0.036 & 18.60$\pm$0.029\\ 
J221924.27+004614.0 & 2.290 & This work & $-$23.25 & 21.64$\pm$0.015 & 21.40$\pm$0.027 & 21.14$\pm$0.065 & 20.63$\pm$0.029 & 20.18$\pm$0.034 & 18.53$\pm$0.027\\ 
J221936.37+002434.12 & 2.852 & SDSS & $-$26.76 & 18.63$\pm$0.002 & 18.48$\pm$0.002 & 18.45$\pm$0.008 & 17.84$\pm$0.004 & 17.35$\pm$0.007 & 16.60$\pm$0.007\\ 
J221937.14+001448.02 & 2.020 & This work & $-$22.36 & 22.08$\pm$0.023 & 21.96$\pm$0.040 & 21.97$\pm$0.182 & 21.63$\pm$0.071 & 21.16$\pm$0.077 & 19.98$\pm$0.082\\ 
J221939.68+000809.7 & 1.960 & This work & $-$21.97 & 22.46$\pm$0.035 & 22.26$\pm$0.056 & 22.36$\pm$0.285 & 21.45$\pm$0.089 & 21.08$\pm$0.072 & 19.84$\pm$0.073\\ 
J221941.99+003631.66 & 1.260 & This work & $-$21.18 & 22.16$\pm$0.025 & 21.86$\pm$0.041 & 21.48$\pm$0.093 & 20.73$\pm$0.040 & 20.59$\pm$0.054 & 18.67$\pm$0.030\\ 
J221942.28+003253.12 & 2.018 & SDSS & $-$24.09 & 20.29$\pm$0.006 & 20.23$\pm$0.010 & 20.03$\pm$0.025 & 19.49$\pm$0.014 & 19.52$\pm$0.025 & 18.44$\pm$0.025\\ 
J221942.46+003415.38 & 1.551 & SDSS & $-$23.64 & 20.14$\pm$0.005 & 19.97$\pm$0.008 & 20.10$\pm$0.027 & 19.82$\pm$0.016 & 19.38$\pm$0.023 & 18.07$\pm$0.019\\ 
J221943.30+004118.38 & 3.133 & SDSS & $-$26.46 & 19.45$\pm$0.003 & 19.03$\pm$0.003 & 19.02$\pm$0.010 & 18.38$\pm$0.006 & 17.85$\pm$0.009 & 16.80$\pm$0.008\\ 
J221945.08+003708.25 & 3.531 & SDSS & $-$27.00 & 19.76$\pm$0.004 & 18.81$\pm$0.003 & 18.76$\pm$0.009 & 18.69$\pm$0.007 & 17.89$\pm$0.009 & 16.75$\pm$0.008\\ 
J221946.66+004340.76 & 1.800 & SDSS & $-$22.75 & 21.48$\pm$0.020 & 21.26$\pm$0.032 & 21.26$\pm$0.081 & 20.40$\pm$0.027 & 20.16$\pm$0.039 & 18.80$\pm$0.034\\ 
J221946.92+000615.78 & 2.020 & This work & $-$22.55 & 21.59$\pm$0.017 & 21.77$\pm$0.036 & 21.36$\pm$0.114 & 20.59$\pm$0.029 & 20.77$\pm$0.057 & 19.28$\pm$0.045\\ 
J221947.10+005526.29 & 2.146 & SDSS & $-$23.28 & 21.37$\pm$0.013 & 21.20$\pm$0.024 & 20.92$\pm$0.064 & 20.41$\pm$0.030 & 19.98$\pm$0.030 & 18.56$\pm$0.028\\ 
J221951.27+004135.30 & 0.656 & VVDS & $-$20.07 & 21.86$\pm$0.018 & 21.21$\pm$0.022 & 21.07$\pm$0.059 & 20.26$\pm$0.020 & 19.93$\pm$0.033 & 18.40$\pm$0.025\\ 
J221952.07+000054.21 & 0.817 & SDSS & $-$21.51 & 20.53$\pm$0.007 & 20.37$\pm$0.010 & 20.13$\pm$0.038 & 19.93$\pm$0.019 & 18.82$\pm$0.016 & 17.45$\pm$0.012\\ 
J221952.12+001933.34 & 0.807 & SDSS & $-$20.15 & 21.86$\pm$0.019 & 21.69$\pm$0.031 & 21.53$\pm$0.120 & 21.11$\pm$0.057 & 20.24$\pm$0.042 & 19.05$\pm$0.042\\ 
J221955.48+004722.78 & 1.770 & This work & $-$23.79 & 20.63$\pm$0.007 & 20.17$\pm$0.009 & 20.20$\pm$0.028 & 19.69$\pm$0.014 & 19.28$\pm$0.019 & 18.22$\pm$0.021\\ 
J221957.22+005521.46 & 1.925 & SDSS & $-$25.48 & 19.06$\pm$0.002 & 18.71$\pm$0.003 & 18.58$\pm$0.008 & 18.21$\pm$0.005 & 17.76$\pm$0.009 & 16.63$\pm$0.007\\ 
J221958.21+003709.33 & 3.089 & SDSS & $-$26.50 & 19.21$\pm$0.003 & 18.95$\pm$0.004 & 18.97$\pm$0.010 & 18.48$\pm$0.006 & 18.00$\pm$0.010 & 17.10$\pm$0.010\\ 
J221958.42+001629.89 & 2.244 & VVDS & $-$23.51 & 21.33$\pm$0.012 & 21.09$\pm$0.018 & 20.89$\pm$0.068 & 20.49$\pm$0.023 & 20.02$\pm$0.033 & 18.63$\pm$0.027\\ 
J222000.42+002137.83 & 1.271 & SDSS & $-$23.16 & 20.19$\pm$0.005 & 19.91$\pm$0.007 & 20.04$\pm$0.031 & 19.19$\pm$0.010 & 19.07$\pm$0.019 & 17.62$\pm$0.014\\ 
J222001.31+000349.28 & 1.422 & SDSS & $-$24.20 & 19.35$\pm$0.003 & 19.17$\pm$0.004 & 19.30$\pm$0.018 & 18.90$\pm$0.008 & 18.44$\pm$0.013 & 17.22$\pm$0.010\\ 
J222001.57+001222.69 & 2.770 & This work & $-$23.37 & 21.98$\pm$0.022 & 21.79$\pm$0.036 & 21.70$\pm$0.156 & 21.45$\pm$0.060 & 20.81$\pm$0.058 & 19.68$\pm$0.064\\ 
J222002.85+004149.77 & 2.330 & SDSS & $-$23.28 & 21.22$\pm$0.018 & 21.42$\pm$0.063 & 21.23$\pm$0.093 & 21.38$\pm$0.061 & 20.45$\pm$0.049 & 18.92$\pm$0.037\\ 
J222007.77+002332.06 & 2.420 & SDSS & $-$24.28 & 21.37$\pm$0.012 & 20.52$\pm$0.011 & 20.17$\pm$0.035 & 19.42$\pm$0.012 & 18.99$\pm$0.018 & 17.67$\pm$0.014\\ 
J222012.53+001051.6 & 1.479 & VVDS & $-$22.96 & 20.70$\pm$0.007 & 20.52$\pm$0.012 & 20.50$\pm$0.051 & 19.98$\pm$0.017 & 19.35$\pm$0.022 & 17.82$\pm$0.015\\ 
J222014.44+001859.16 & 2.490 & This work & $-$23.83 & 20.97$\pm$0.009 & 21.05$\pm$0.017 & 20.62$\pm$0.052 & 20.16$\pm$0.023 & 19.52$\pm$0.025 & 18.22$\pm$0.022\\ 
J222014.59+004238.23 & 1.900 & This work & $-$22.45 & 21.98$\pm$0.021 & 21.70$\pm$0.034 & 21.69$\pm$0.106 & 20.83$\pm$0.034 & 20.25$\pm$0.041 & 18.95$\pm$0.041\\ 
J222015.45+002601.46 & 2.243 & SDSS & $-$22.86 & 22.06$\pm$0.024 & 21.75$\pm$0.036 & 21.48$\pm$0.091 & 20.76$\pm$0.040 & 20.36$\pm$0.044 & 18.74$\pm$0.033\\ 
J222020.40+001047.71 & 1.489 & SDSS & $-$23.93 & 19.95$\pm$0.004 & 19.57$\pm$0.005 & 19.61$\pm$0.023 & 19.07$\pm$0.012 & 18.54$\pm$0.013 & 17.53$\pm$0.013\\ 
J222028.54+000531.63 & 2.748 & SDSS & $-$24.67 & 21.08$\pm$0.010 & 20.47$\pm$0.011 & 20.15$\pm$0.037 & 19.26$\pm$0.011 & 18.99$\pm$0.017 & 17.78$\pm$0.015\\ 
J222029.53+004401.32 & 0.621 & SDSS & $-$22.62 & 18.56$\pm$0.002 & 18.52$\pm$0.002 & 18.61$\pm$0.007 & 17.57$\pm$0.005 & 17.59$\pm$0.008 & 16.09$\pm$0.005\\ 
J222032.50+002537.66 & 4.196 & SDSS & $-$26.34 & 21.90$\pm$0.021 & 19.91$\pm$0.007 & 19.87$\pm$0.021 & 19.13$\pm$0.010 & 18.75$\pm$0.015 & 17.41$\pm$0.012\\ 
J222034.37+005723.4 & 2.170 & This work & $-$22.16 & 22.40$\pm$0.033 & 22.34$\pm$0.069 & 21.95$\pm$0.164 & 21.36$\pm$0.049 & 21.14$\pm$0.069 & 19.21$\pm$0.042\\ 
J222035.99+005339.3 & 1.426 & VVDS & $-$21.55 & 22.86$\pm$0.049 & 21.83$\pm$0.042 & 21.71$\pm$0.128 & 20.86$\pm$0.035 & 20.12$\pm$0.034 & 18.49$\pm$0.025\\ 
J222037.15+001426.72 & 1.880 & This work & $-$21.84 & 22.39$\pm$0.030 & 22.28$\pm$0.052 & 21.95$\pm$0.179 & 20.97$\pm$0.039 & 20.88$\pm$0.062 & 19.52$\pm$0.060\\ 
J222040.96+000531.41 & 2.500 & SDSS & $-$23.85 & 21.10$\pm$0.011 & 21.04$\pm$0.019 & 20.83$\pm$0.069 & 20.47$\pm$0.031 & 20.17$\pm$0.037 & 18.48$\pm$0.025\\ 
J222043.88+002354.31 & 0.521 & SDSS & $-$20.46 & 20.35$\pm$0.006 & 20.21$\pm$0.009 & 20.09$\pm$0.033 & 19.41$\pm$0.012 & 18.99$\pm$0.018 & 17.31$\pm$0.011\\ 
J222047.75+000853.08 & 0.985 & SDSS & $-$21.92 & 21.02$\pm$0.010 & 20.45$\pm$0.011 & 19.94$\pm$0.031 & 19.30$\pm$0.011 & 18.56$\pm$0.014 & 17.06$\pm$0.009\\ 
J222050.60+005948.51 & 2.601 & SDSS & $-$25.48 & 20.46$\pm$0.006 & 19.52$\pm$0.006 & 19.37$\pm$0.017 & 18.68$\pm$0.007 & 18.42$\pm$0.012 & 17.29$\pm$0.010\\ 
J222052.10+001024.92 & 2.463 & SDSS & $-$23.68 & 21.41$\pm$0.014 & 21.17$\pm$0.021 & 20.85$\pm$0.070 & 20.35$\pm$0.025 & 20.00$\pm$0.033 & 18.47$\pm$0.025\\ 
J222052.76+004917.61 & 2.200 & SDSS & $-$23.96 & 20.68$\pm$0.007 & 20.59$\pm$0.013 & 20.34$\pm$0.032 & 19.81$\pm$0.017 & 19.48$\pm$0.022 & 17.92$\pm$0.016\\ 
J222055.88+005219.41 & 2.604 & SDSS & $-$24.83 & 20.14$\pm$0.005 & 20.16$\pm$0.960 & 19.92$\pm$0.025 & 19.62$\pm$0.012 & 19.26$\pm$0.020 & 18.05$\pm$0.018\\ 
J222057.44+000329.98 & 2.260 & SDSS & $-$25.85 & 19.15$\pm$0.003 & 18.77$\pm$0.003 & 18.49$\pm$0.009 & 17.86$\pm$0.004 & 17.56$\pm$0.008 & 16.26$\pm$0.006\\ 
J222057.76+005105.25 & 1.013 & SDSS & $-$22.30 & 20.29$\pm$0.006 & 20.15$\pm$0.009 & 20.06$\pm$0.028 & 19.45$\pm$0.014 & 18.93$\pm$0.016 & 17.47$\pm$0.012\\ 
J222058.98+005917.08 & 2.644 & SDSS & $-$25.86 & 19.36$\pm$0.003 & 19.18$\pm$0.007 & 19.02$\pm$0.013 & 18.45$\pm$0.007 & 18.03$\pm$0.010 & 17.13$\pm$0.009\\ 
J222059.51+003840.95 & 2.820 & VVDS & $-$23.57 & 22.80$\pm$0.044 & 21.64$\pm$0.032 & 21.35$\pm$0.078 & 20.53$\pm$0.030 & 20.03$\pm$0.035 & 18.65$\pm$0.032\\ 
J222100.33+005320.43 & 1.285 & VVDS & $-$21.56 & 22.01$\pm$0.023 & 21.54$\pm$0.032 & 21.36$\pm$0.093 & 20.57$\pm$0.027 & 20.27$\pm$0.038 & 18.41$\pm$0.026\\ 
J222100.87+000950.97 & 1.354 & VVDS & $-$21.46 & 22.08$\pm$0.025 & 21.78$\pm$0.036 & 21.47$\pm$0.122 & 20.55$\pm$0.036 & 20.41$\pm$0.044 & 18.88$\pm$0.035\\ 
J222103.42+005836.4 & 4.220 & This work & $-$25.05 & 23.31$\pm$0.075 & 21.21$\pm$0.025 & 21.06$\pm$0.074 & 20.74$\pm$0.037 & 20.16$\pm$0.035 & 19.14$\pm$0.040\\ 
J222103.64+002203.46 & 1.187 & VVDS & $-$21.85 & 21.26$\pm$0.011 & 21.04$\pm$0.017 & 20.81$\pm$0.063 & 20.21$\pm$0.021 & 19.83$\pm$0.029 & 18.33$\pm$0.022\\ 
J222103.80+004820.95 & 0.750 & This work & $-$21.04 & 20.59$\pm$0.007 & 20.61$\pm$0.014 & 20.39$\pm$0.034 & 19.99$\pm$0.023 & 19.34$\pm$0.021 & 17.56$\pm$0.014\\ 
J222105.68+003101.83 & 3.160 & This work & $-$23.67 & 22.32$\pm$0.032 & 21.84$\pm$0.042 & 21.69$\pm$0.111 & 21.08$\pm$0.047 & 20.50$\pm$0.044 & 18.63$\pm$0.027\\ 
J222110.30+002740.1 & 1.280 & This work & $-$21.38 & 22.21$\pm$0.028 & 21.71$\pm$0.036 & 21.73$\pm$0.114 & 20.84$\pm$0.041 & 20.50$\pm$0.046 & 19.03$\pm$0.038\\ 
J222112.97+010115.88 & 2.006 & SDSS & $-$24.05 & 20.43$\pm$0.007 & 20.25$\pm$0.033 & 20.14$\pm$0.151 & 19.32$\pm$0.047 & 19.62$\pm$0.025 & 18.36$\pm$0.022\\ 
J222118.57+001144.66 & 2.157 & SDSS & $-$25.07 & 19.55$\pm$0.003 & 19.43$\pm$0.005 & 19.23$\pm$0.017 & 19.36$\pm$0.010 & 18.95$\pm$0.017 & 17.33$\pm$0.011\\ 
J222121.18+001247.36 & 2.240 & This work & $-$22.92 & 22.25$\pm$0.027 & 21.67$\pm$0.030 & 21.44$\pm$0.114 & 20.71$\pm$0.034 & 20.66$\pm$0.055 & 19.18$\pm$0.044\\ 
J222126.96+001451.09 & 2.660 & This work & $-$23.97 & 21.45$\pm$0.013 & 21.08$\pm$0.018 & 20.92$\pm$0.070 & 20.19$\pm$0.018 & 20.07$\pm$0.036 & 18.88$\pm$0.035\\ 
J222128.67+001443.58 & 1.682 & VVDS & $-$23.75 & 20.34$\pm$0.010 & 20.08$\pm$0.026 & 20.13$\pm$0.065 & 19.53$\pm$0.014 & 19.11$\pm$0.019 & 17.90$\pm$0.016\\ 
J222128.71+004455.86 & 2.134 & SDSS & $-$25.04 & 19.62$\pm$0.003 & 19.43$\pm$0.005 & 19.19$\pm$0.012 & 18.71$\pm$0.007 & 18.53$\pm$0.013 & 17.18$\pm$0.011\\ 
J222129.87+000430.07 & 2.476 & SDSS & $-$25.78 & 19.15$\pm$0.002 & 19.08$\pm$0.004 & 18.88$\pm$0.012 & 18.30$\pm$0.005 & 18.02$\pm$0.010 & 16.83$\pm$0.008\\ 
J222132.56+010005.43 & 3.833 & VVDS & $-$24.84 & 22.17$\pm$0.027 & 21.18$\pm$0.025 & 21.19$\pm$0.086 & 20.45$\pm$0.040 & 20.14$\pm$0.034 & 19.12$\pm$0.039\\ 
J222133.06+004040.36 & 2.197 & SDSS & $-$23.88 & 20.83$\pm$0.008 & 20.67$\pm$0.014 & 20.41$\pm$0.034 & 20.34$\pm$0.051 & 19.70$\pm$0.026 & 18.24$\pm$0.020\\ 
J222136.96+001144.23 & 1.420 & SDSS & $-$23.96 & 19.55$\pm$0.003 & 19.41$\pm$0.005 & 19.51$\pm$0.020 & 19.24$\pm$0.009 & 18.63$\pm$0.014 & 17.11$\pm$0.009\\ 
J222138.01+001559.6 & 2.060 & This work & $-$23.22 & 21.26$\pm$0.011 & 21.15$\pm$0.019 & 21.00$\pm$0.076 & 20.82$\pm$0.041 & 20.17$\pm$0.038 & 19.02$\pm$0.039\\ 
J222143.21+002550.25 & 0.990 & This work & $-$20.71 & 22.64$\pm$0.041 & 21.68$\pm$0.035 & 21.12$\pm$0.067 & 20.23$\pm$0.022 & 19.47$\pm$0.023 & 17.99$\pm$0.017\\ 
J222143.61+002456.41 & 1.119 & SDSS & $-$24.22 & 18.57$\pm$0.002 & 18.51$\pm$0.002 & 18.54$\pm$0.009 & 17.98$\pm$0.006 & 17.92$\pm$0.010 & 16.67$\pm$0.007\\ 
J222146.13+003745.24 & 1.750 & This work & $-$23.27 & 20.82$\pm$0.008 & 20.66$\pm$0.014 & 20.59$\pm$0.040 & 20.36$\pm$0.042 & 19.61$\pm$0.024 & 18.35$\pm$0.022\\ 
J222146.71+002303.91 & 2.195 & SDSS & $-$23.29 & 21.41$\pm$0.013 & 21.25$\pm$0.022 & 21.08$\pm$0.082 & 20.65$\pm$0.041 & 20.41$\pm$0.043 & 18.77$\pm$0.031\\ 
J222150.92+001345.44 & 0.784 & SDSS & $-$21.17 & 20.58$\pm$0.007 & 20.59$\pm$0.012 & 20.26$\pm$0.039 & 19.57$\pm$0.029 & 18.89$\pm$0.017 & 17.47$\pm$0.012
\enddata
\end{deluxetable*}

\begin{deluxetable*}{ c c c c c c| c c c c }
\centering
\tablecaption{Binned QLF \& Parametric QLF}
\tablewidth{0pt}
\tablehead{
  \colhead{redshift range} &
  \colhead{$<z>$} &
  \colhead{$M_{1450}$ bin} &
  \colhead{$<M_{1450}>$} &
  \colhead{$log(\Phi)$} &
  \colhead{$\sigma_{\phi}$\tablenotemark{a}} &
  \colhead{$log(\Phi^{*})$\tablenotemark{c}} &
  \colhead{$\sigma_{log(\Phi^{*})}$} &
  \colhead{$\alpha$} &
  \colhead{$\sigma_{\alpha}$}
  }
\startdata
   0.5$<z<$1.0 & 0.760 & $-$23.5 & $-$23.410 & $-$6.174 & 0.474 & -6.090 & 0.256 & -1.561 & 0.337\\
                        &           & $-$22.5 & $-$22.588 & $-$6.169 & 0.479 &            &           &            &           \\
                        &           & $-$21.5 & $-$21.363 & $-$5.662 & 0.902 &            &           &            &           \\
                        &           & $-$20.5 & $-$20.422 & $-$5.457 & 1.935 &            &           &            &           \\
                        &           & & & & & & & & \\
  1.0$<z<$1.5  & 1.303 & $-$25.823\tablenotemark{b} & $-$25.823 & $-$6.687 & 0.206 & -6.104 & 0.204 & -1.468 & 0.288\\
                        &          & $-$24.5 & $-$24.212 & $-$6.393 & 0.286 &            &           &            &           \\
                        &          & $-$23.5 & $-$23.658 & $-$5.907 & 0.506 &            &           &            &           \\
                        &          & $-$22.5 & $-$22.539 & $-$5.889 & 0.527 &            &           &            &           \\
                        &          & $-$21.5 & $-$21.566 & $-$5.481 & 1.331 &            &           &            &           \\
                        &           & & & & & & & & \\                     
  1.5$<z<$2.0 & 1.751 & $-$25.485 & $-$25.485 & $-$6.772 & 0.169 & -5.889 & 0.478 & -1.028 & 0.631\\
                        &          & $-$23.5 & $-$23.491 & $-$5.859 & 0.533 &            &           &            &           \\
                        &          & $-$22.5 & $-$22.627 & $-$5.948 & 0.518 &            &           &            &           \\
                        &          & $-$21.5 & $-$21.904 & $-$5.339 & 2.629 &            &           &            &           \\
                        &           & & & & & & & & \\
  2.0$<z<$2.5 & 2.235 & $-$25.5 & $-$25.436 & $-$6.194 & 0.320 & -6.165 & 0.194 & -1.853 & 0.305\\
                        &          & $-$24.5 & $-$24.202 & $-$6.094 & 0.361 &            &           &            &           \\
                        &          & $-$23.5 & $-$23.511 & $-$5.652 & 0.687 &            &           &            &           \\
                        &          & $-$22.5 & $-$22.558 & $-$5.323 & 1.766 &            &           &            &           \\
                        &           & & & & & & & & \\
  2.5$<z<$3.0 & 2.700 & $-$26.765 & $-$26.765 & $-$6.792 & 0.162 & -6.141 & 0.227 & -1.114 & 0.470\\
                        &          & $-$25.5 & $-$25.669 & $-$6.501 & 0.223 &            &           &            &           \\
                        &          & $-$24.5 & $-$24.813 & $-$6.191 & 0.322 &            &           &            &           \\
                        &          & $-$23.5 & $-$23.69 & $-$6.061 & 0.444 &            &           &            &           \\
                        &           & & & & & & & & \\
  3.0$<z<$3.5 & 3.119 & $-$26.5 & $-$26.483 & $-$6.478 & 0.235 & -6.388 & 0.676 & -1.453 & 0.968\\
                        &          & $-$24.5 & $-$24.457 & $-$6.156 & 0.349 &            &           &            &           \\
                        &          & $-$23.5 & $-$23.607 & $-$5.98 & 0.667 &            &           &            &           \\
                        &           & & & & & & & & \\
  3.5$<z<$4.0 & 3.682 & $-$26.998 & $-$26.998 & $-$6.771 & 0.170 &   --      &     --     &    --     & --   \\
                        &          & $-$24.835 & $-$24.835 & $-$6.739 & 0.183 &            &           &            &    \\
                        &           & & & & & & & & \\
  4.0$<z<$4.5 & 4.208 & $-$26.342 & $-$26.342 & $-$6.738 & 0.183 &     --     &  --     &   --     &  --  \\
                        &          & $-$25.050 & $-$25.050 & $-$6.532 & 0.294 &            &           &            &           
 \enddata
\tablenotetext{a}{$\sigma_{\phi}$ is in units of $\rm 10^{-6} Mpc^{-3} mag^{-1}$s.}
\tablenotetext{b}{For magnitude bins including only one quasar, we use the $M_{1450}$ of this quasar.}
\tablenotetext{c}{The best fits at each redshift bin. Since we fix the bright end slope $\beta$ and break magnitude $M^{*}_{1450}$ to \cite{croom09} and \cite{ross13} at different redshift range, here we only list the best fits and uncertainties of $log(\Phi^{*})$ and $\alpha$.}
\end{deluxetable*}

\acknowledgments
J. Y. and X.-B. W. thank the supports by the NSFC grant No.11373008 and 11533001, the Strategic Priority Research Program ''The Emergence of Cosmological Structures'' of the Chinese Academy of Sciences, Grant No. XDB09000000, the National Key Basic Research Program of China 2014CB845700, and from the Ministry of Science and Technology of China under grant 2016YFA0400703. D.-Z.L., S.Y., and Z.-H.F. acknowledge the support from NSFC of China under grants 11333001,11173001, and 11033005, and from the Strategic Priority Research Program '' The Emergence of Cosmological Structure'' of the Chinese Academy of Science, grant no. XDB09000000. X.F. and I.D.M. acknowledge the support from US NSF grant AST 15-15115 and NASA ADAP grant NNX17AF28G. H.-Y.S. acknowledges the support from TR33 project ''The Dark Universe'' funded by the DFG. This research uses data obtained through the Telescope Access Program (TAP), which has been funded by the Strategic Priority Research Program "The Emergence of Cosmological Structures" (Grant No. XDB09000000), National Astronomical Observatories, Chinese Academy of Sciences, and the Special Fund for Astronomy from the Ministry of Finance in China. We acknowledge the use of the CFHT and the MMT 6.5 m telescopes. This work was partially supported by the Open Project Program of the Key Laboratory of Optical Astronomy, National Astronomical Observatories, Chinese Academy of Sciences. 

We acknowledge the use of SDSS photometric data. Funding for SDSS-III has been provided by the Alfred P. Sloan Foundation, the Participating Institutions, the National Science Foundation, and the U.S. Department of Energy Office of Science. The SDSS-III Web site is http://www.sdss3.org/. SDSS-III is managed by the Astrophysical Research Consortium for the Participating Institutions of the SDSS-III Collaboration including the University of Arizona, the Brazilian Participation Group, Brookhaven National Laboratory, University of Cambridge, Carnegie Mellon University, University of Florida, the French Participation Group, the German Participation Group, Harvard University, the Instituto de Astrofisica de Canarias, the Michigan State/Notre Dame/JINA Participation Group, Johns Hopkins University, Lawrence Berkeley National Laboratory, Max Planck Institute for Astrophysics, Max Planck Institute for Extraterrestrial Physics, New Mexico State University, New York University, Ohio State University, Pennsylvania State University, University of Portsmouth, Princeton University, the Spanish Participation Group, University of Tokyo, University of Utah, Vanderbilt University, University of Virginia, University of Washington, and Yale University. We acknowledge the use of the UKIDSS data and Hubble Source Catalog \footnote{Based on observations made with the NASA/ESA Hubble Space Telescope, and obtained from the Hubble Legacy Archive, which is a collaboration between the Space Telescope Science Institute (STScI/NASA), the Space Telescope European Coordinating Facility (ST-ECF/ESAC/ESA) and the Canadian Astronomy Data Centre (CADC/NRC/CSA).}.

{\it Facilities:} \facility{Sloan (SDSS)}, \facility{UKIRT/WFCAM}, \facility{CFHT/WIRCam}, \facility{MMT/Hectospec}.

\end{document}